\documentclass[10pt, aps,prb,amsmath,superscriptaddress,amssymb,longbibliography,twocolumn]{revtex4-2}
\usepackage{amsmath}
\usepackage{graphicx}
\graphicspath{ {images/}}
\usepackage{dcolumn}
\usepackage{bm}
\usepackage[	
citecolor=magenta, colorlinks, linkcolor=magenta, urlcolor=magenta]{hyperref}
\usepackage{braket}
\usepackage{appendix}
\usepackage{wasysym}
\usepackage{slashed}
\usepackage[table, svgnames, dvipsnames]{xcolor}
\usepackage{textcomp}
\usepackage{amsmath}
\usepackage{array, multirow, bigdelim, makecell, booktabs} 
\usepackage{soul}
\usepackage{graphicx}
\usepackage{subfigure}
\usepackage{ragged2e}

\newcommand{\up}{\uparrow}
\newcommand{\dn}{\downarrow}
\usepackage{setspace}

\begin{document}


\title{Role of topotactic hydrogen in Superconductivity of Infinite-layer Nickelate NdNiO\texorpdfstring{$_2$}{2}: A first-principles and variational Monte Carlo study}

\author{Manoj Gupta}
\email{Equal contribution first author}
 \affiliation{Department of Condensed Matter and Materials Physics,
S. N. Bose National Centre for Basic Sciences, Kolkata 700106, India}
\author{Arun Kumar Maurya}
\email{Equal contribution first author}
\affiliation{Department of Condensed Matter and Materials Physics,
S. N. Bose National Centre for Basic Sciences, Kolkata 700106, India}
\author{Amal Medhi}
\affiliation{School of Physics, IISER Thiruvananthapuram, Thiruvananthapuram - 695551,
Kerala, India.}
\author{Tanusri Saha Dasgupta}
\email{t.sahadasgupta@gmail.com}
\affiliation{Department of Condensed Matter and Materials Physics,
S. N. Bose National Centre for Basic Sciences, Kolkata 700106, India}
\date{\today}

\begin{abstract}

Employing combination of first-principles calculations, low-energy model construction, and variational Monte Carlo solution of the ab-initio derived Hubbard model, we study 
the effect of hydrogenation in the electronic structure and superconducting 
properties of infinite-layer nickelate, NdNiO$_2$. We find that the introduction of hydrogen at the apical oxygen
vacancy position strongly influences the Wannier function corresponding to the effective interstitial orbital at the Ni site bound to the hydrogen. This results in the near disappearance of the electron pocket at the $k_z$ = $\pi$
Fermi surface, keeping that of $k_z$ = 0 unchanged, compared to the dehydrogenated case. The two-band model thus remains valid even in the presence of H. The calculated superconducting
order parameters both in absence and presence of H, show a two-hump superconductivity arising the two overlapping domes, one arising from $d_{x^{2}-y^{2}}$ and another arising 
from interstitial orbital degree of freedom. Hydrogenation strengthens the latter, marginally affecting the former.

\end{abstract}
\maketitle

\section{Introduction}
A long-standing quest since the discovery of superconductivity is finding
of candidate materials exhibiting superconductivity at elevated temperature.
Even after more than 30 years of research in the cuprate family of superconductors, understanding its physical origin remains an open question~\cite{Bednorz1986,anderson1987resonating,RevModPhys.66.763,RevModPhys.72.969,annurevLouis,siegrist1988crystal,balestrino2002evidence,dicastro2015direct}. 
Alternative to cuprates, Fe superconductors~\cite{doi:10.1021/ja063355c,doi:10.1021/ja800073m,doi:10.1143/JPSJ.79.102001,RevModPhys.83.1589}, nickelates
were discussed. Among the alternatives, nickelates are most attractive as
Ni may host same $d^9$ (Ni$^{1+}$) configuration as Cu $d^9$ (Cu$^{2+}$).
Although a number of attempts were made~\cite{chaloupka2008orbital,hansmann2009turning} including superlattices of LaNiO$_3$ and trilayer reduced La$_3$Ni$_2$O$_6$ and La$_4$Ni$_3$O$_8$~\cite{doi:10.1021/ja063031o,doi:10.1021/ic701480v,PhysRevB.84.180411}, finally in 2019,  superconductivity in the nickelate family was observed in infinite-layer hole-doped compound, (Nd,Sr)NiO$_2$, starting the era of nickelate superconductors~\cite{nature2019nickelate,nomura2022superconductivity,osada2021nickelate,Norman2020NickelAge}. Much of the excitement after the discovery was in understanding the similarities and differences between cuprates and nickelates.
While the undoped cuprates are antiferromagnetic with strong insulating properties~\cite{lee2006doping}, undoped nickelates have been reported to be nonmagnetic with metallic or bad metallic behavior~\cite{PhysRevResearch.3.L042015}. The x-ray absorption spectroscopic (XAS) and resonant inelastic x-ray scattering (RIXS) data~\cite{PhysRevB.104.L220505} confirm that nickelates
are more in the Mott-Hubbard regime, as opposed to charge-transfer regime of cuprates. Furthermore, the theoretical investigations~\cite{nomura2019formation,lee2004infinite,botana2020similarities,sakakibara2020model}  as well as analysis of spectroscopic data point~\cite{Hepting2020} to a two band scenario, contributed by Ni $d_{x^{2}-y^{2}}$ and a delocalized
interstitial $s$-like orbital (IS) which is a hybrid of different orbital degrees of freedom involving rare-earth and empty Ni 4$s$ orbital~\cite{PhysRevB.102.100501,PhysRevB.102.220502}.

Parallel to this, unlike cuprates, for which superconductivity is widely reported in both bulk and powder samples by many different experimental groups, nickelate superconductivity has so far only been observed 
in thin-film samples grown on certain substrates and only by 
a limited number of experimental groups. One of the main challenges in the synthesis of nickelates lies in achieving the ultra-low valence state of Ni$^{1+}$. This requires a strong topotactic reduction process, such as using the metal-hydride CaH$_2$ agent~\cite{yamamoto2013hydride,Matsumoto2019SmFeAsO}, which converts perovskite nickelates to the infinite-layer structure by removing apical oxygen atoms. This may naturally lead to the insertion of hydrogen~\cite{GAINZA2023101724} in 
nickelates during the growth process. Detection of H is challenging due to its light weight and insensitivity to most characterisation techniques.
Though the signal of hydrogen was not observed at all in the nickelates using the advanced transmission electron microscope (TEM), a recent study~\cite{Ding2023HydrogenNickelates} using ultra-sensitive secondary-ion mass spectroscopy (SIMS) to characterise the elemental distributions in (Nd,Sr)NiO$_2$, supports 
the presence of H in the sample. The same study~\cite{Ding2023HydrogenNickelates} also indicates superconductivity
is observed for a rather narrow H concentration window of 0.22 $\leq$ x $\leq$ 0.28.
While this generated a lot of interest, it remains a debated issue, with counter reports of the coexistence of low hydrogen concentrations with superconductivity~\cite{Balakrishnan2024HydrogenNickelates, Gonzalez2024HydrogenNickelates}. The role of the capping layer in the thin-film geometry also remains controversial~\cite{10.1063/5.0005103,Krieger2022Nickelate,Krieger_2023,10.1063/5.0197304}.

Given the complexity and details of differences in different experimental setups, it is prudent to have an independent theoretical study understanding the influence of H in superconductivity---a crucial step for optimising these materials for superconducting applications. 
We investigate this issue within the framework
of first-principles density functional theory (DFT) calculations and variational Monte Carlo (VMC) study of a DFT-derived two-band Hubbard Hamiltonian. 
While the SIMS finds the existence of hydrogen in the sample, the exact locations remain unknown. Also, H as a unique impurity, can be either in +1 or -1 charge state. The previous DFT calculations~\cite{Si2020} revealed
that the H atoms are not randomly distributed; rather, they prefer to 
sit at the apical oxygen vacancy (AOV) positions, with large in-plane 
H-H separation. This results in the formation of H–Ni–H linear chain 
in the out-of-plane direction, and thus recovers the local octahedral structure. Based on this H configuration, it is found that H impurities in nickelates exhibit a negative -1 charge state~\cite{Ding2023HydrogenNickelates} at the interstitial site, in contrast to some of the iron-based superconductors~\cite{CUI201811}. In our study, we thus
considered H at AOV positions with 25$\%$ concentration, which is within
the suggested window of 0.22 $\leq$ x $\leq$ 0.28 and compared its superconducting properties with the dehydrogenated case. Our study reveals that within the two orbital
description of the compounds, while the hydrogenation leaves the Ni-$d_{x^{2}-y^{2}}$
Wannier function more or less unchanged, it strongly affects the local IS Wannier function for the Ni site closely bound to the H at the AOV site (Ni$_{AOV}$). This changes the Fermi surface topology at $k_z$ = $\pi$ plane, leaving that at $k_z$ = 0 plane more or
less unchanged, compared to the dehydrogenated case. This in turn reduces the $d_{x^{2}-y^{2}}$-IS
local hopping interaction at the Ni$_{AOV}$ site, at the
cost of significant enhancement of the out-of-plane IS-IS hopping. The $d_{x^{2}-y^{2}}$-IS hopping, as well as IS-IS hopping at the Ni site closest to Ni$_{AOV}$, also get significantly influenced. The VMC calculation on the two-band Hubbard model
based on DFT-derived tight-binding model in the Ni-$d_{x^{2}-y^{2}}$-IS Wannier function basis, shows a two-dome superconductivity, as seen in the experiments~\cite{Zeng2020PhaseDiagram,Chow2022InfiniteLayer,Li2020SuperconductingDome, Osada2022FieldEffect}
for both dehydrogenated and hydrogenated compounds. Our study further
highlights the orbital-selective nature of superconductivity with the first dome arising from IS, the second one being contributed by $d_{x^{2}-y^{2}}$. The $d_{x^{2}-y^{2}}$ contributed dome shows only a marginal reduction upon hydrogenation, while the IS contributed dome shows an appreciable enhancement. We thus conclude, 
H, if present in the sample, strengthens the IS-derived superconductivity, marginally affecting $d_{x^{2}-y^{2}}$ derived superconductivity.


\section{Computational Details}
The first-principles DFT calculations were performed using a plane-wave basis set. These calculations employed the projector augmented-wave (PAW) method~\cite{PhysRevLett.77.3865}, as implemented in the Vienna Ab-initio Simulation Package (VASP)~\cite{PhysRevB.50.17953,TACKETT2001348,10.1063/1.1926272}. The convergence of total energies and forces was ensured by using a plane-wave energy cutoff of 600 eV and Brillouin zone (BZ) sampling with a $6 \times 6 \times 6$ Monkhorst-Pack grid. During structural relaxation, the ions were allowed to move until the atomic forces were reduced below 0.0001 eV/\AA. The Perdew-Burke-Ernzerhof (PBE) generalised gradient approximation (GGA)~\cite{PhysRevLett.77.3865} was used to approximate the exchange-correlation functional. To construct a low-energy Hamiltonian, we employed the maximally localised Wannier function (MLWF) technique, implemented in the WANNIER90 code~\cite{MOSTOFI20142309}, for both hydrogenated and dehydrogenated systems. The initial guess for the orbital basis set consisted of Ni-$d_{x^2-y^2}$ and Ni-$s$ orbitals.  
For band unfolding, we evaluated the weight factor of each supercell band at every $k$-point along high-symmetry lines by projecting onto the primitive cell using self-consistent field (SCF) calculations in the plane-wave basis. The unfolded band structures were then plotted using the VASPKIT code~\cite{WANG2021108033}.
 
The VMC~\cite{SHIBA1987264,PhysRevB.70.054504,doi:10.1143/JPSJ.57.2482, PhysRevB.76.235122,doi:10.1143/JPSJ.77.114701,Sorella2013,Sarder_2022} calculations were carried in a 
Hubbard model constructed using the DFT-derived Wannier basis. The simulations are
done on a cubic lattice of size $8\times 8\times 6$ with $L=384$ sites. With each site
containing two orbitals, there are  a total of $768$ orbital degrees of freedom. Thus, we have
a very large system size which makes the simulations highly challenging. 
For calculations of observables, we used $\sim 5\times10^3$ thermalization Monte Carlo  
sweeps followed by $\sim 1\times10^5$ sweeps for measurement. In the variational wave function, 
which is a Bardeen–Cooper–Schrieffer (BCS) mean-field state correlated by a long range Jastrow projector,
we have a total of 184 variational parameters which includes 180 Jastrow factors apart from
various gap parameters in the mean-field part. The wave function was optimized using 
stochastic reconfiguration method~\cite{PhysRevB.64.024512} which works well even for large number of variational parameters. 
The system size dependency was checked by considering a few different lattice sizes.

\section{Crystal Structure}
\begin{figure}[htb]
    \centering
    \includegraphics[width=0.8\linewidth]{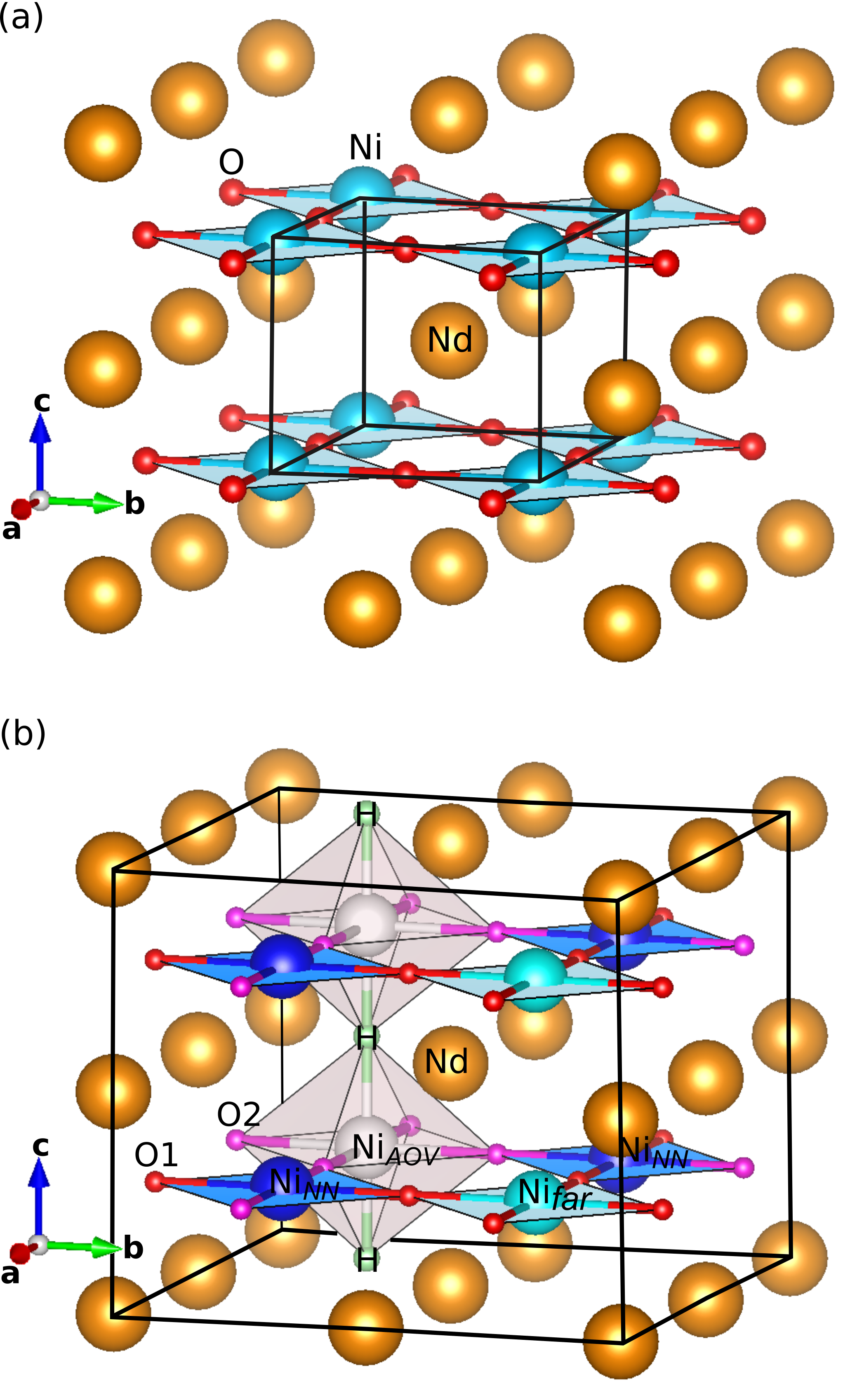}
    \caption{(a) The crystal structure of dehyrogenated NdNiO$_2$ featuring square planner O coordination of Ni. The atoms are shown as balls. The
    box highlights the tetragonal unit cell. (b) The crystal structure of 
    25$\%$ hyrogenated compound (NdNiO$_2$H$_{0.25}$). Inequivalent atoms
    are colored differently. Ni$_{AOV}$, Ni$_{NN}$/Ni$_{far}$ are 
    in octahedral and  tetrahedral coordination, respectively. The box marks the 2$\times$2$\times$2 supercell used for calculation.}
    \label{fig:structure}
\end{figure}

Dehydrogenated NdNiO$_2$ crystallizes in a tetragonal lattice with space group P4/mmm (No. 123), characterized by lattice parameters $a = b \neq c$ with $a = 3.92$ \AA and $c = 3.28$ \AA~\cite{HAYWARD2003839}. The structure possesses a fourfold rotational symmetry axis ($C_4$) along the $z$-axis, three mirror planes ($xy$, $yz$, and $xz$), and inversion symmetry. The unit cell consists of one Ni, one Nd, and three O atoms, with Wyckoff positions and site symmetries given by Ni: $1a$, site symmetry $4/mmm$; Nd: $1d$, site symmetry $4/mmm$; O: $2f$, site symmetry $mmm$. 
The oxygen atoms form a square planar coordination around Ni, creating a NiO$_4$ square lattice, with Nd atoms sandwiched between these Ni-O planes, as shown in 
Fig.~\ref{fig:structure}(a). Due to this square planar coordination, the symmetry of the Ni $d$-orbitals is reduced from spherical to $D_{4h}$ symmetry.

\begin{figure*}[htb]
    \centering
    \includegraphics[width=0.85\linewidth]{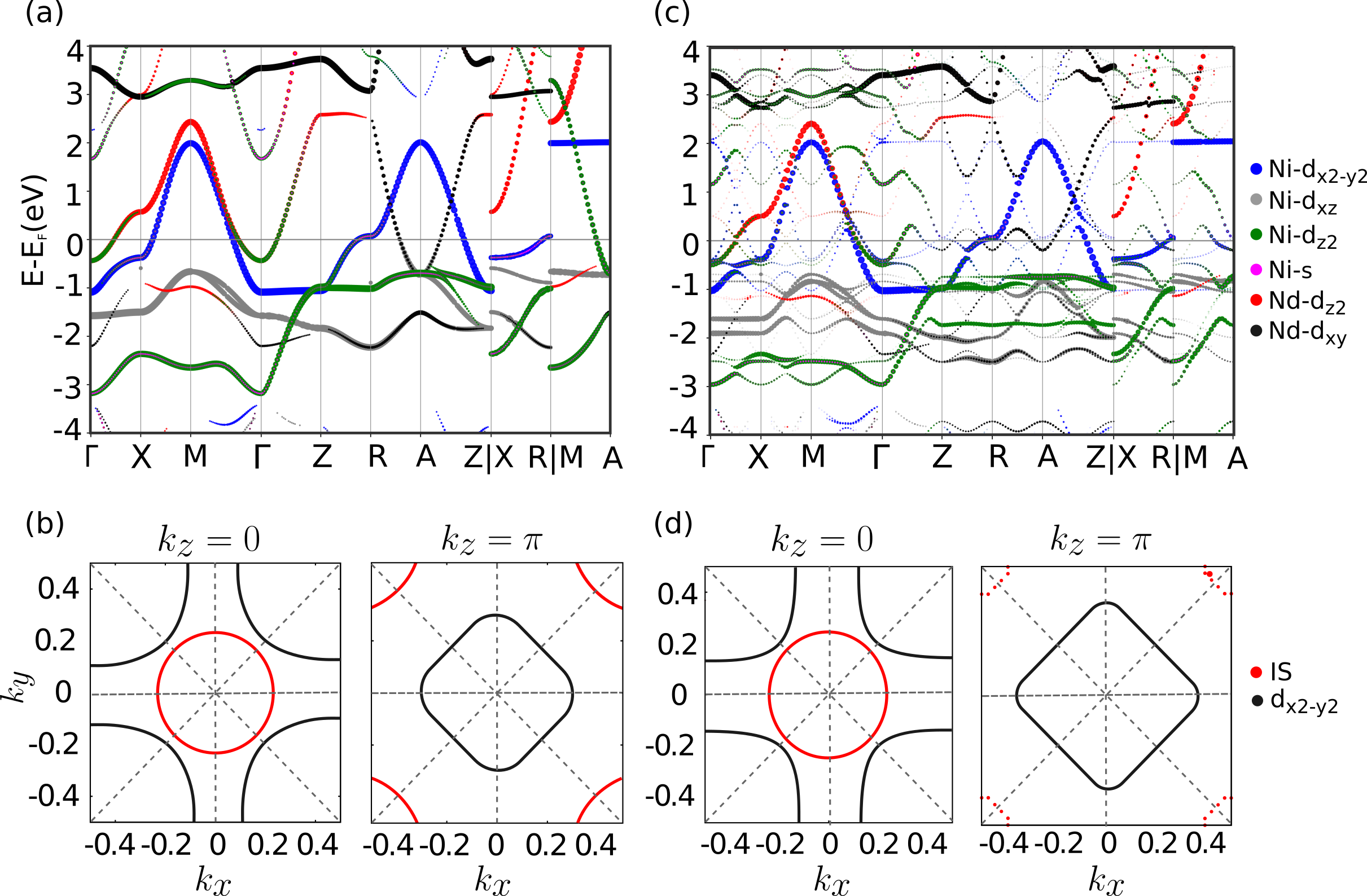}
    \caption{(a) The GGA bandstructure of dehydrogenated NdNiO$_2$ compound,
    plotted along the high symmetry paths of the tetragonal BZ, $\Gamma$= (0,0,0),
    $X$ = ($\pi$,0,0), $M$= ($\pi$,$\pi$,0), $Z$=(0,0,$\pi$), $R$ = ($\pi$, 0, $\pi$), $A$= ($\pi$, $\pi$, $\pi$). The zero of the energy is fixed GGA Fermi energy and the bands was projected to Ni($d_{x^2-y^2}$, $d_{xy}$, $d_{z^2}$, $d_{xz}$, $d_{yz}$) and Nd($d_{z^2}$, $d_{xy}$)
    orbital characters. (b) The corresponding FS at $k_z$ = 0 and $k_z$ = $\pi$ planes. The contributions of $d_{x^2-y^2}$ (black) and IS (red) 
    orbitals are marked. (c) Same as (a), but plotted for hydrogenated compound (NdNiO$_2$H$_{0.25}$). The bands are unfolded in the tetragonal BZ
    of NdNiO$_2$ to compare with the bandstructure of dehydrogenated compound.
    (d) Same as (b), plotted for NdNiO$_2$H$_{0.25}$.}
    \label{fig:band-unfolding}
\end{figure*}

As mentioned above, we considered 25$\%$ hydrogenated compound (NdNiO$_2$H$_{0.25}$) with the hydrogen impurity at the AOV position. Based on this, we constructed a $2\times2\times2$  supercell containing 8 Ni sites and introduced two hydrogen atoms at AOV 
positions $(0.25, 0.25, 0)$ and $(0.25, 0.25, 0.5)$, forming a Ni-H-Ni chain in the out-of-plane direction. The structure was optimized
theoretically keeping the symmetry intact.
Introduction of hydrogen impurities in the cell lowers the site symmetries: Ni atoms get split into three inequivalent sites at Wyckoff positions $1b$, $1d$, and $2e$; oxygen atoms split into two inequivalent sites with symmetries $4m$ ($m2m$) and $4o$ ($m2m$). All Nd atoms remain symmetry-equivalent, with
$4f$ Wyckoff positions. Out of the three inequivalent Ni atoms, as shown in Fig.~\ref{fig:structure}(b),
Ni$_{AOV}$ at Wyckoff position $1d$ gets strongly bound to H atoms forming NiO$_4$H$_2$ octahedral coordination, as compared to NiO$_4$ square planar coordination in the other two inequivalent Ni sites, as well as that in the dehydrogenated situation. Among the two other Ni sites, Ni$_{NN}$ at Wyckoff position $2e$ is the nearest neighbour to Ni$_{AOV}$, 
and thus feels effect of H in the next level, Ni$_{far}$ at Wyckoff position $1b$ being the farthest from Ni$_{AOV}$.
\section{First-Principles Results}

\subsection{Basic Electronic Structure}
The square planar geometry of NiO$_4$ in dehydrogenated nickelates leads to an energy-level splitting of the Ni $d$-orbitals in the following order, $d_{x^2-y^2} > d_{xy} > d_{z^2} > (d_{xz}, d_{yz})$. The larger $d$-$p$ charge transfer
energy for nickelates compared to cuprates, places the Ni-$d$ levels higher up,
facilitating its hybridisation with nominally empty rare-earth 5$d$ states.
The well-studied low-energy GGA band structure of NdNiO$_2$ \cite{nomura2019formation,lee2004infinite,botana2020similarities}, thus features two bands crossing the Fermi energy, as shown in  Fig.~\ref{fig:band-unfolding}(a),
plotted along the high symmetry paths of the tetragonal Brillouin zone (BZ).
Out of the bands, one is derived from the Ni-$d_{x^2-y^2}$ orbital and another is from the IS orbital that emerges from hybridization of Ni-$s$, Ni-$d_{z^2}$, Ni-$d_{xz}$, Nd-$d_{z^2}$, and Nd-$d_{xy}$. These two bands give rise to two Fermi surface (FS) sheets. At $k_z = 0$ plane, the Ni-$d_{x^2-y^2}$ band contributes to a zone-centred circular sheet, whereas the Ni-IS orbital gives rise to a zone-cornered FS, as shown in Fig.~\ref{fig:band-unfolding}(b). This gets reversed at $k_z = \pi$, where the Ni-$d_{x^2 - y^2}$ band forms a zone-cornered FS, while the Ni-IS orbital generates a zone-centred one. 

The band structure of hydrogenated NdNiO$_2$, unfolded in the tetragonal BZ
of the unit cell of dehydrogenated nickelates for comparison, is shown in
Fig.~\ref{fig:band-unfolding}(c). Upon hydrogenation, significant modification
happens in the band structure in $k_z = \pi$ plane, leaving the band structure in $k_z = 0$ more or
less unchanged. The anionic $s$-type charge density of impurity H atoms, strongly interact with out-of-plane $d$ orbitals ($d_{xz}$, $d_{yz}$, $d_{z^2}$) of Ni$_{AOV}$, leading to bonding–antibonding splitting (cf. Appendix~\ref{ref:appnxa}). Notably, the orbital occupancy of the $d_{z^2}$ orbital decreases by approximately $13\%$ upon hydrogen doping.  
This makes the energy position of the IS band
shifted up in $k_z$ = $\pi$ plane, thus making the IS-derived electron pocket  
at A point nearly disappear. This is reflected in the FS plot at $k_z = \pi$,
where $d_{x^2 - y^2}$ FS expands into a larger area and the FS associated with 
IS almost disappears (cf. Fig.~\ref{fig:band-unfolding}(d)). However, the effect of IS remains imprinted in the FS topology of $k_z$ = 0 plane, even after hydrogenation, stressing upon
the requirement of a multi-orbital description of the problem even in the presence of H.

\subsection{Wannier-functions and Low-energy Hamiltonian}

\begin{figure}[htp]
    \centering
    \includegraphics[width=1.0\linewidth]{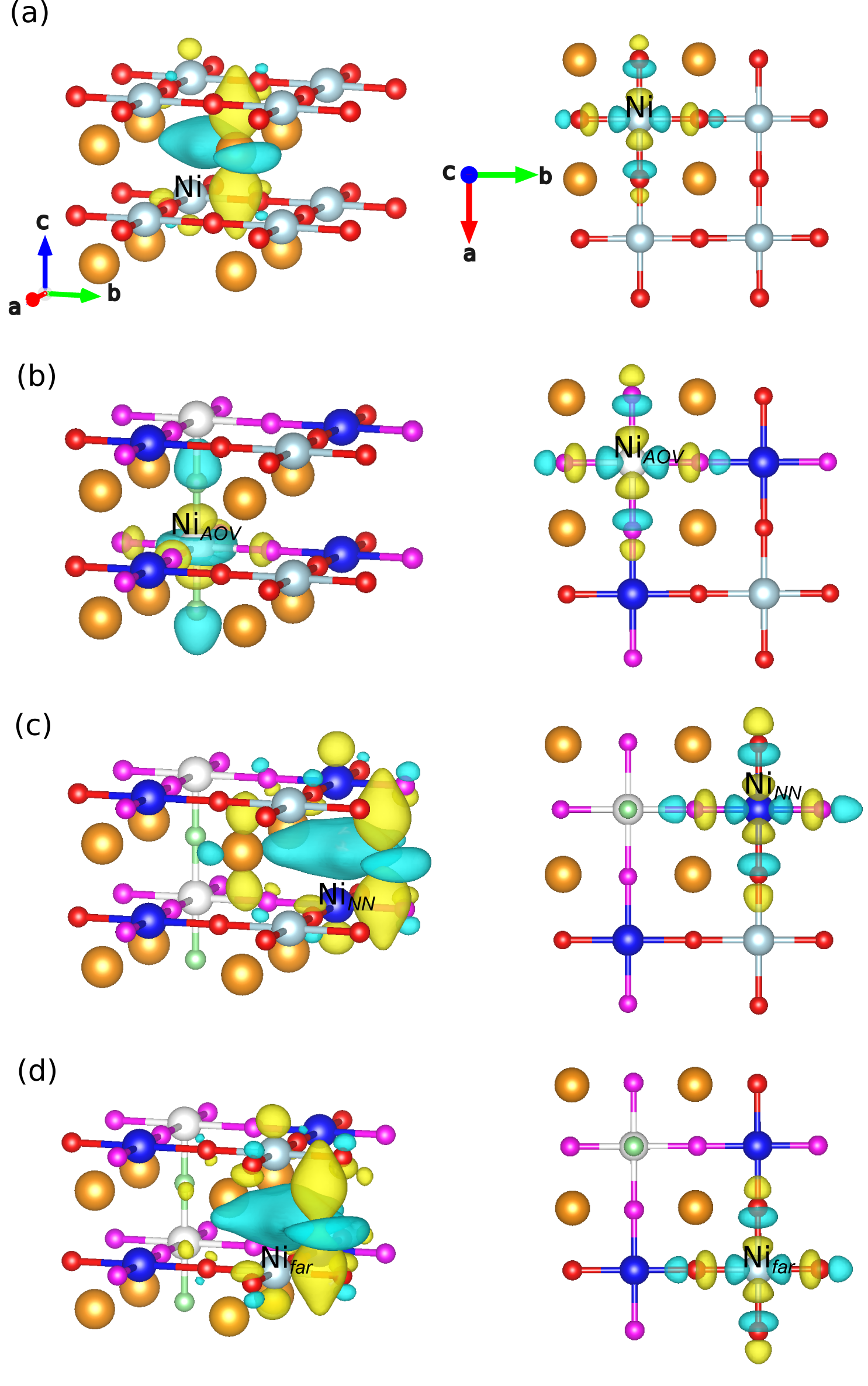}
  \caption{Low-energy Wannier functions for $d_{x^2-y^2}$ (right) and IS (left) orbitals corresponding to the two-orbital downfolded bands. Plotted are
  the constant value surfaces with positive and negative iso-surfaces colored
  as cyan and yellow, respectively. Shown are for (a) Ni in NdNiO$_2$ compound,
  (b) Ni$_{AOV}$  in NdNiO$_2$H$_{0.25}$ compound, (c) Ni$_{NN}$ in NdNiO$_2$H$_{0.25}$ compound, and (d) Ni$_{far}$ in NdNiO$_2$H$_{0.25}$ compound.}
  \label{fig:wannier_orbitals}
\end{figure}

Based on the above, we constructed a low-energy 
Hamiltonian for the dehydrogenated and hydrogenated NdNiO$_2$. For this purpose,  two degrees of freedom per Ni, $d_{x^2-y^2}$ 
and a second degree of freedom(s) to capture the IS orbital were retained in the basis, and the remaining degrees of freedom were integrated out, starting from GGA self-consistent bandstructure, as formulated in the Wannier90 code~\cite{MOSTOFI20142309}.
For the dehydrogenated situation, all Ni atoms are equivalent, resulting
in two distinct Wannier functions, shown in Fig.~\ref{fig:wannier_orbitals}(a), one shaped as antibonding Ni $d_{x^2-y^2}$-O $p_x/p_y$ pd$\sigma$ 
and another IS, formed from hybridization of Ni-$s$, Ni-$d_{z^2}$,
Ni-$d_{xz}$, Nd-$d_{z^2}$, and Nd-d$_{xy}$. For the hydrogenated situation,
there are six distinct Wannier functions, two for Ni$_{AOV}$, two for Ni$_{NN}$
and two for Ni$_{far}$. While Ni $d_{x^2-y^2}$-O $p_x/p_y$ Wannier functions
for all three Ni's are similar to that of the dehydrogenated case, the IS Wannier
functions are not (cf. Fig.~\ref{fig:wannier_orbitals}(b)-(d)). Specifically, the IS Wannier function of Ni$_{AOV}$ shows
a highly localised nature with limited in-plane spatial extension and strongly bonded to H $s$ in the out-of-plane direction. The IS Wannier function for
Ni$_{NN}$ shows a somewhat more extended nature in-plane, while Ni$_{far}$ IS Wannier function is very similar to that of the dehydrogenated case, implying rather local influence of the impurity H.

\begin{figure}[htb]
    \centering
    \includegraphics[width=0.95\linewidth]{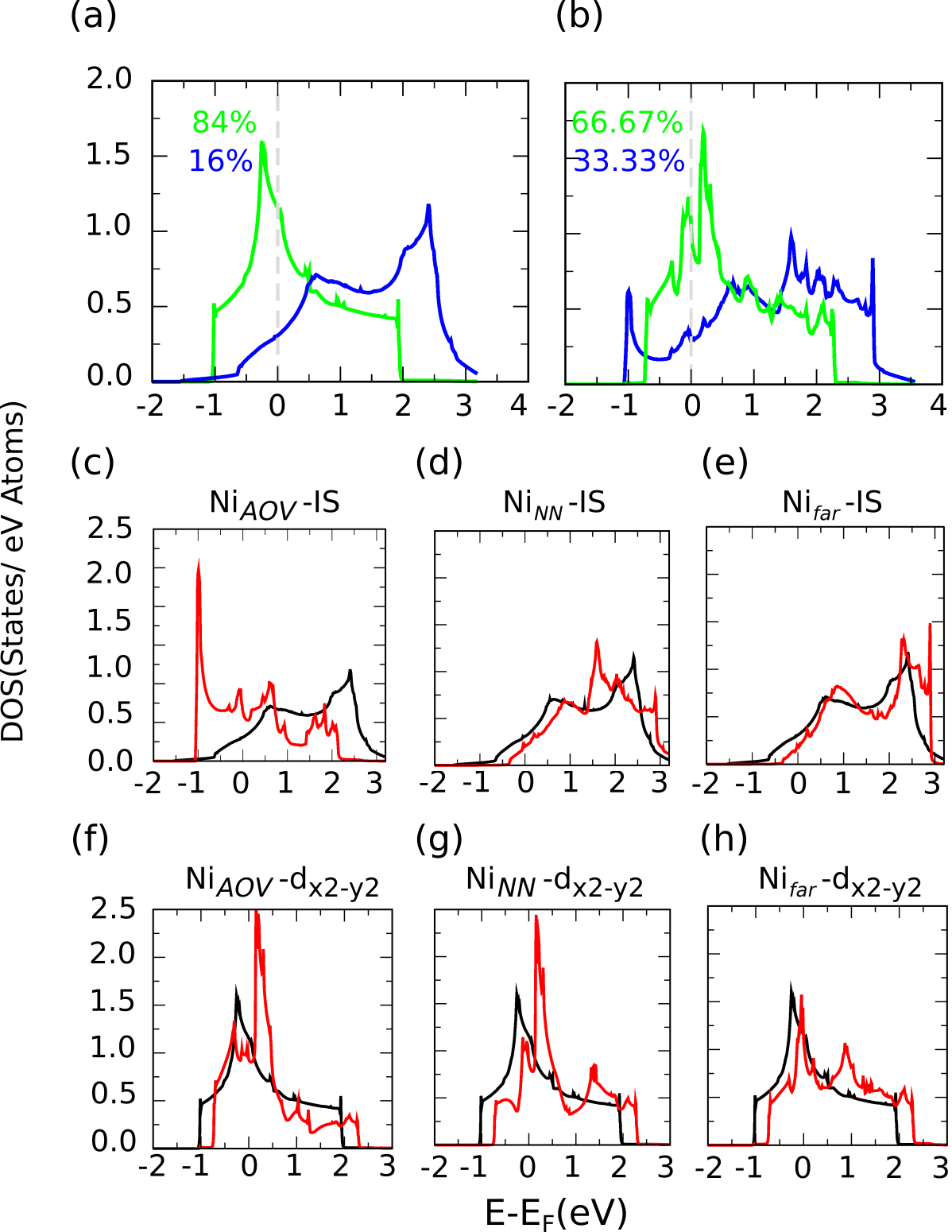}
    \caption{Density of states of the low-energy two band model of (a)  NdNiO$_2$ 
    and (b) NdNiO$_2$H$_{0.25}$. The density of states are projected to $d_{x^2-y^2}$ (green) and IS (blue) degrees of freedom.  The percentage contribution
    of each degree of freedom until Fermi energy is marked. (c)-(h) 
    IS and $d_{x^2-y^2}$ projected DOS for Ni$_{AOV}$, Ni$_{NN}$ and Ni$_{far}$ (red) in NdNiO$_2$H$_{0.25}$ compared to that  in NdNiO$_2$ (black).}
    \label{fig:dos}
\end{figure}

The density of states (DOS) of the low-energy Hamiltonian, projected to $d_{x^2-y^2}$ and IS Wannier functions of the dehydrogenated and hydrogenated compounds
are shown in Fig.~\ref{fig:dos}(a) and (b), respectively. In the dehydrogenated case, as found previously~\cite{PhysRevB.102.100501}, the IS hole dopes $d_{x^2-y^2}$, with
84$\%$ of the electron density at $d_{x^2-y^2}$ and rest at IS.
Upon hydrogenation, transfer of electrons happens between IS and $d_{x^2 - y^2}$,
with 66.7$\%$ of the electron density at $d_{x^2 - y^2}$ and 33.3 $\%$ at IS.
Fig.~\ref{fig:dos}(c)-(h) show the comparison of IS-projected and $d_{x^2 - y^2}$-projected DOS of Ni$_{AOV}$, Ni$_{NN}$ and Ni$_{far}$ to that of dehydrogenated projected DOS. As expected, the maximum deviation is observed
for the IS-projected DOS of Ni$_{AOV}$, exhibiting the signature of strong binding to H $s$. The  $d_{x^2 - y^2}$-projected DOS of Ni$_{AOV}$ as well
as Ni$_{NN}$ shows increased prominence of van Hove singularity, signifying enhanced two dimensionality of $d_{x^2 - y^2}$ Wannier function.

The hopping interactions between the Wannier functions centred at
Ni$_{AOV}$, Ni$_{NN}$ and Ni$_{far}$ can be obtained from the real space
description of the low-energy Hamiltonian in Ni$_{AOV}$ ($d_{x^2 - y^2}$, IS),
Ni$_{NN}$ ($d_{x^2 - y^2}$, IS) and Ni$_{far}$ ($d_{x^2 - y^2}$, IS) Wannier function basis.  
\begin{figure*}[htp]
    \centering
    \includegraphics[width=0.9\linewidth]{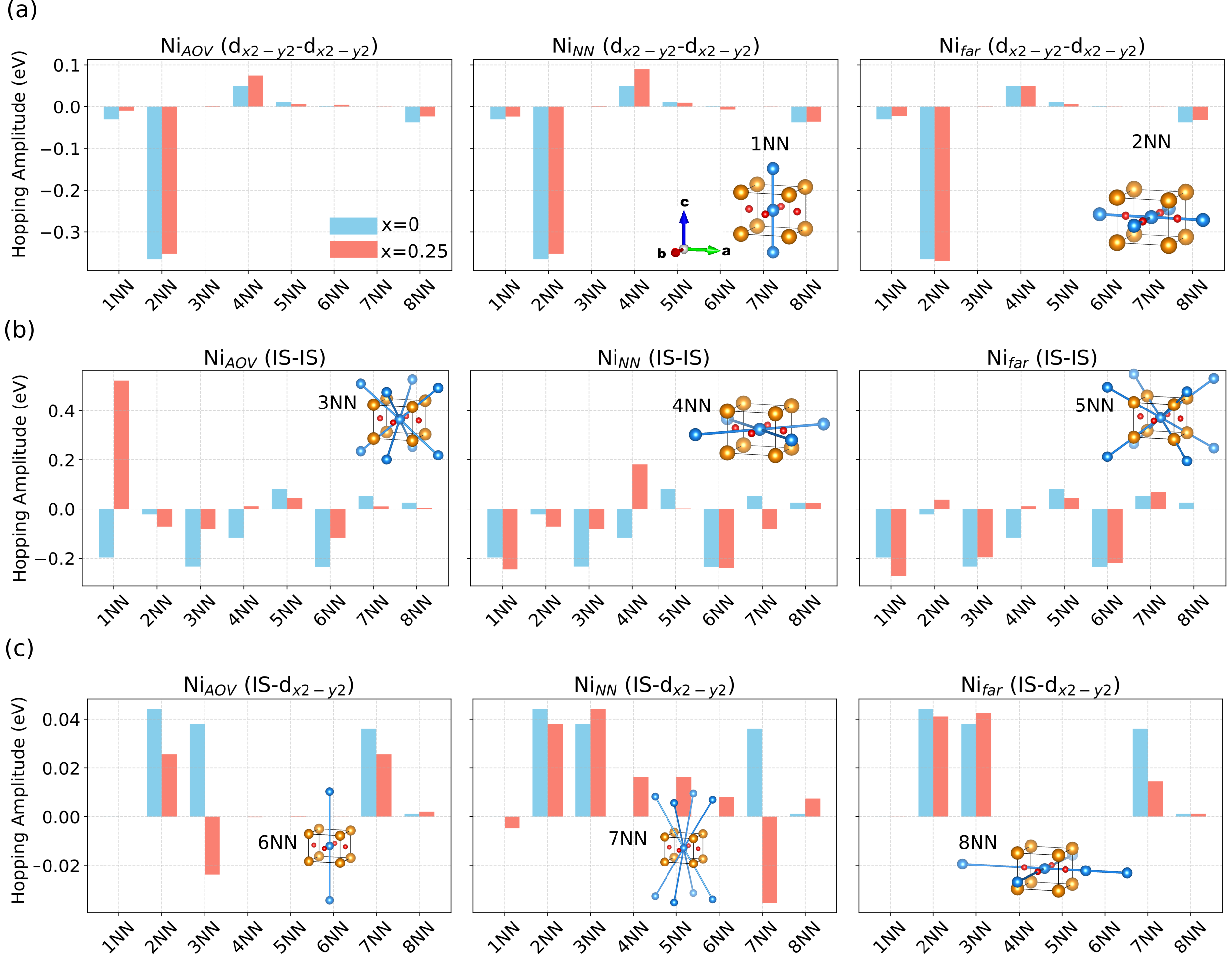}
    \caption{Hopping interactions between Ni-$d_{x^2-y^2}$ and IS Wannier functions in NdNiO$_2$ ($x$ = 0) and NdNiO$_2$H$_{0.25}$ ($x$ = 0.25) plotted as a function of increasing separation of neighbours. The neighbours are shown
    as insets, with Nd, Ni and O atoms colored as orange, blue and red, respectively. Shown are (a) $d_{x^2-y^2}$-$d_{x^2-y^2}$ interactions.
    (b) IS-IS interactions and (c) IS-$d_{x^2-y^2}$ interactions.}
    \label{fig:hopping-parameters}
\end{figure*}
The hopping interactions are found to be non-negligible up to the 8th NN,
which includes both out-of-plane and in-plane hoppings, as shown in the insets of
Fig.~\ref{fig:hopping-parameters}. In the main panels of the figure, hoppings between $d_{x^2-y^2}$ Wannier functions, $d_{x^2-y^2}$ and IS Wannier functions
and IS-IS Wannier functions are shown for Ni$_{AOV}$, Ni$_{NN}$ and Ni$_{far}$ sites in comparison to that of the dehydrogenated situation
(cf. Fig.~\ref{fig:hopping-parameters}(a)-(c)) . $d_{x^2 - y^2}$-$d_{x^2 - y^2}$ hoppings 
are found to be rather similar to that of the dehydrogenated situation,
in accordance with the conclusion drawn from Wannier function plots. A marked
increase in 1NN IS-IS hopping for Ni$_{AOV}$ is observed compared to the dehydrogenated case, due to the directed nature of Ni$_{AOV}$ IS Wannier function.
The $d_{x^2 - y^2}$-IS hopping of Ni$_{AOV}$ on the other hand, shows a suppression, in agreement with DFT results in Ref.~\cite{Ding2023HydrogenNickelates}. The  $d_{x^2-y^2}$-IS hopping of Ni$_{NN}$, shows some modifications, while Ni$_{far}$ remains
more or less the same as the dehydrogenated situation. The combined effect of these modifications in the single-particle Hamiltonian upon hydrogenation
is expected to influence the superconducting properties, which will be taken up
next within the framework of VMC.

\section{Variational Monte Carlo Results} 
To study the superconducting properties of the dehydrogenated and hydrogenated
compounds, we employed the variational Monte Carlo technique. To the best of our knowledge, the VMC technique has not been employed so far in the study of
superconductivity in nickelates, although it has been extensively used
for the study of superconductivity in cuprates~\cite{YANAGISAWA20021379,PhysRevResearch.3.033157,PhysRevX.13.041036}.
Furthermore, although there exists theoretical study of dehydrogenated 
nickelate superconductivity using techniques like DFT+random-phase approximation~\cite{PhysRevB.102.100501,PhysRevB.102.220502}, 
DFT+ dynamical vertex approximation (D$\Gamma$A)~\cite{DiCataldo2024,hausoel2025}, linearized GW+dynamical mean field theory (DMFT)~\cite{Kang2023},
the fluctuation-exchange approximation~\cite{sakakibara2020model}, as well as from
strong-coupling starting points~\cite{PhysRevB.101.020501,PhysRevResearch.2.023112,Chang_2020}, to the best of our knowledge, there does not exist similar theoretical exercise on the superconductivity of hydrogenated nickelate.

\begin{figure*}[!htp]
\centering
\includegraphics[width=0.8\linewidth]{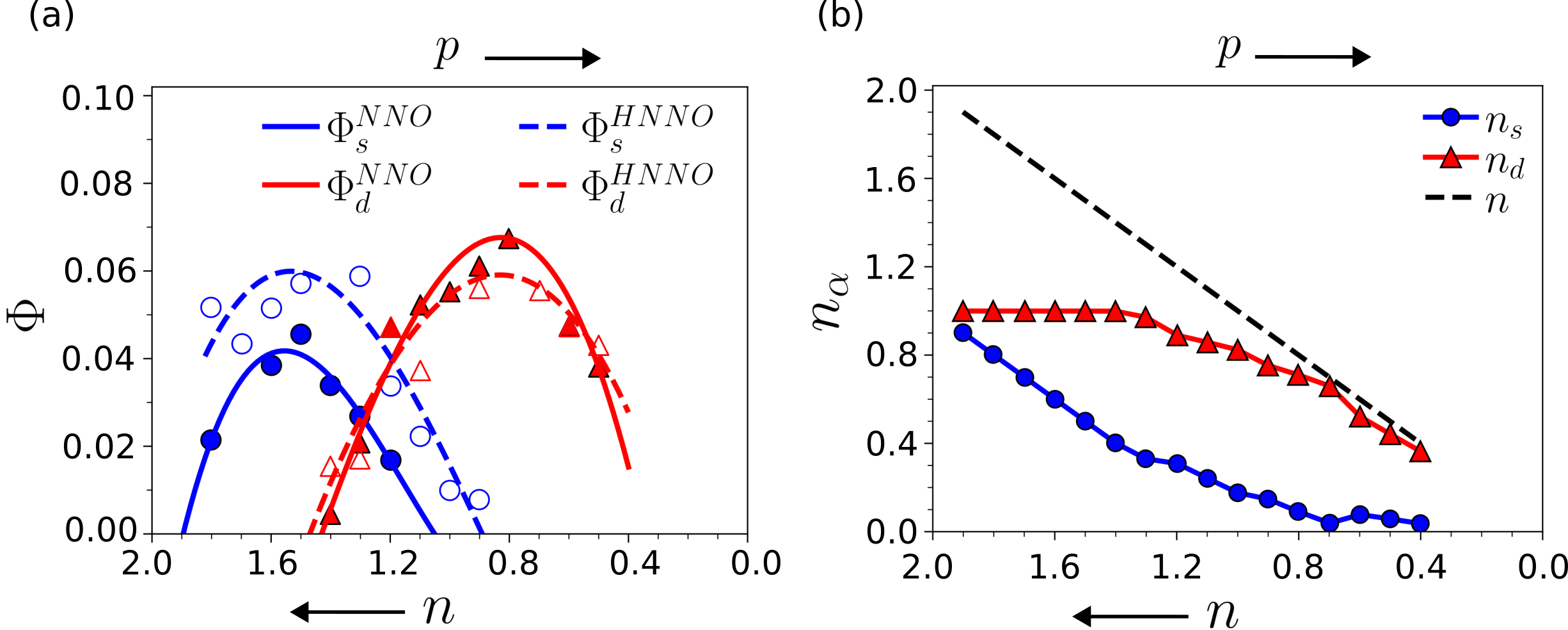}
\caption{(a) Superconducting order parameter, $\Phi_{\alpha}$ ($\alpha$ = $d$; $s$) for hydrogenated NdNiO$_2$, compared to dehydrogenated NdNiO$_2$ 
as a function of electron density $n$. The  open and closed symbols indicate
VMC data corresponding to hydrogenated and dehydrogenated NdNiO$_2$, respectively. 
The solid and dashed lines are guide to eye corresponding to data points for 
hydrogenated and dehydrogenated  NdNiO$_2$, respectively. (b) The orbital resolved
electron density, $n_{\alpha}$ ($\alpha$ = $d$; $s$) plotted as function
of total electron density. The dashed line marks the fall of total electron density.
The Hubbard interactions are fixed at $U_{d}=3$ eV, $\frac{U_s}{U_d}$ = 0.5 and $\frac{U_{sd}}{U_d}$ = 0.2.}
\label{fig:vmc-results}
\end{figure*}

For this purpose, we constructed a two-band Hubbard model, with its one-electron 
part given by the tight-binding Hamiltonian derived from the DFT Wannier functions,
described above (cf. Fig.~\ref{fig:hopping-parameters} and related discussion),
\begin{align}
{\cal H} = &\sum_{\substack{i j \sigma \\ \alpha \beta}} t^{\alpha\beta}_{ij} \Bigl(c^\dag_{i\alpha \sigma}c_{j\beta \sigma} + \text{hc} \Bigr) - \sum_{i\alpha \sigma} \epsilon_{\alpha} n_{i\alpha\sigma} \nonumber \\
&+ \sum_{i\alpha} U_{\alpha} n_{i\alpha\up}n_{i\alpha\dn} + \sum_{i\sigma\sigma'}
U_{sd} n_{is\sigma}n_{id\sigma'}.
\label{eq:Hubbard2B}
\end{align}
Here $i$, $j$ are the site indices, $\alpha$, $\beta$ (=$s$ (IS),\; 
$d$ ($d_{x^2 - y^2}$)) are the orbital indices, and $\sigma$, $\sigma'$ ($=\up, \; \dn$) are the spin indices.
$c^{\dag}_{i\alpha\sigma}$ ($c_{i\alpha\sigma}$) is the electron creation (annihilation) operator, and the number operator is represented by $n_{i\alpha\sigma}=c^{\dag}_{i\alpha\sigma}c_{i\alpha\sigma}$.  
The first term represents the hopping of electrons between sites, which includes both intra- and inter-orbital hoppings. The second term represents the onsite orbital energy. 
The last two terms represent intra- and inter-orbital onsite Hubbard interaction. 
For numerical ease, the spin-exchange and pair-hopping terms that arise in a general multi-orbital Hubbard Hamiltonian have not been considered.
An important set of parameters that needs to be chosen in the above Hamiltonian is
\{$U_d$, $U_s$ and $U_{sd}$\}. Given the localised character of $d_{x^2 - y^2}$ Wannier function and rather delocalized nature of IS, we take
$U_s$ and $U_{sd}$ fractions of $U_d$. 
We have checked results by considering a few different values of
$U_d$ and varying ratios of $\frac{U_s}{U_d}$ and $\frac{U_{sd}}{U_d}$.
The results presented in Fig.~\ref{fig:vmc-results} are obtained with
$\frac{U_s}{U_d}$ = 0.5, $\frac{U_{sd}}{U_d}$ = 0.2 and $U_d$ = 3 eV.
Variation of $U_d$ over 1-2 eV, and the $\frac{U_s}{U_d}$, 
$\frac{U_{sd}}{U_d}$ ratios by 0.2-0.3, the qualitative trend 
was found to remain the same, though the detailed nature was 
found to vary.
To solve the Hamiltonian in Eq.~\ref{eq:Hubbard2B}, we consider 
a variational ground state of the following form,
\begin{align}
\ket{\Psi_{var}} = {\cal P}_J \ket{BCS},
\label{eq:VarWF}
\end{align}
where $\ket{BCS}$ is the BCS ground state (with fixed particle number) of the following 
pairing mean-field  Hamiltonian,
\begin{align}
{\cal H}_{MF} =& \sum_{\substack{i j \sigma \\ \alpha \beta}} \bigl(t_{ij}^{\alpha \beta} c^\dag_{i\alpha\sigma}c_{j\beta\sigma}+\text{hc} \bigr) + \sum_{i\alpha\sigma}(\epsilon_{\alpha} - \mu_{\alpha}) \ c^\dag_{i\alpha\sigma}c_{i\alpha\sigma} \nonumber \\
& -\sum_{ij\alpha} \bigl[\Delta^{\alpha}_{ij}\bigl( c_{i\alpha\up}^{\dag}c_{j\alpha\dn}^{\dag}-c_{i\alpha\dn}^{\dag}c_{j\alpha\up}^{\dag} \bigr) + \text{hc} \bigr],
\label{eq:MF}
\end{align}
where $\mu_{\alpha}$ is the Hartree shift or chemical potential, which is taken as a variational
parameter. The last term describes the superconducting pairing of electrons. 
Here we consider only intra-orbital pairing between nearest neighbour sites. Motivated
by the past theoretical findings~\cite{PhysRevB.102.100501,PhysRevB.104.144504} of two-gap superconductivity in nickelates, supported by the scanning tunneling microscopy (STM) study~\cite{Gu2020NatComm}, and Hall resistivity 
data~\cite{PhysRevLett.125.027001}, we take the pairing  symmetry to be $s$-wave for the IS orbital and $d$-wave for the $d_{x^2-y^2}$ orbital. Thus,
\begin{align}
\Delta^{s}_{i,i+\delta} =& \Delta^s_{sc} \;\text{for}\; \delta = \hat{x},\;\hat{y},\;\hat{z} \nonumber \\
\Delta^{d}_{i,i+\delta} =& \Delta^d_{sc}  \text{for}\ \delta = \hat{x},\;-\Delta^d_{sc} \text{for}\ \delta = \hat{y},\; 0 \;\text{for}\; \delta = \hat{z} 
\end{align}
where the quantities $\Delta^s_{sc}$, $\Delta^d_{sc}$ are variational parameters.
${\cal P}_{J}=\text{exp}\left(-\frac{1}{2} \sum_{ij\alpha\beta} v_{ij}^{\alpha \beta} 
n_{i\alpha} n_{j\beta} \right)$ is the long range density-density Jastrow projection operator~\cite{PhysRevLett.94.026406,doi:10.1143/JPSJ.75.114706,PhysRevB.98.075117, PhysRevB.110.125125}, which is  an extension of the original onsite Gutzwiller projection operator~\cite{PhysRevLett.10.159,doi:10.1143/JPSJ.56.1490}.
The parameters $v_{ij}^{\alpha \beta}$ are the Jastrow variational parameters, which are functions of the distance between two sites, i.e., $v_{ij}^{\alpha \beta} \equiv v_{ij}^{\alpha \beta} \bigl(|\mathbf{r}_{i} - \mathbf{r}_{j}|\bigr)$.
The wave function is optimised by minimising the variational energy with respect to 
the variational parameters using the stochastic reconfiguration method~\cite{PhysRevB.64.024512}. 

We study the superconducting (SC) properties of the two systems as a function of the 
electron density, $n=N_e/L$, where $N_e$ is the total number of electrons and $L$ is the number of sites in the lattice. We compute the SC pair-pair correlation function
given by,
$F_{\alpha}(\mathbf{r}_i - \mathbf{r}_j) = \langle B^{\dagger}_{i,i+\delta;\alpha} B_{j,j+\delta';\alpha}\rangle$, where $B^{\dagger}_{i,i+\delta;\alpha} = 
\frac{1}{\sqrt{2}}\left(c^{\dagger}_{i\alpha\up}c^{\dagger}_{i+\delta\alpha\dn}-
c^{\dagger}_{i\alpha\dn}c^{\dagger}_{i+\delta\alpha\up}\right)$, $\delta=\hat{x},
\;\hat{y},\;\hat{z}$.
The SC order parameter $\Phi$ is then obtained as,
\begin{align}
\Phi^2_{\alpha} = \lim_{|\mathbf{r}_i - \mathbf{r}_j| \to \infty} F_{\alpha}(\mathbf{r}_i - \mathbf{r}_j).
\end{align}

The results for the SC order parameters $\Phi_{d}^{NNO}$, $\Phi_{s}^{NNO}$ for the dehydrogenated compound and $\Phi_{d}^{HNNO}$, $\Phi_{s}^{HNNO}$ for the hydrogenated compound are shown in Fig.~\ref{fig:vmc-results}(a). As the figure shows, the SC order parameters for both the systems show a double-dome structure as a function of particle density, $n$, the corresponding hole concentration being $p=2-n$, measured from half-filling of two-band Hamiltonian.
Near $n=2$, the $\Phi_\alpha$'s are vanishingly small. 
As holes are introduced, the $\Phi_s$ value corresponding to $s$-wave SC correlations in the IS orbital starts to increase. It peaks at $n\sim 1.6$ and then decreases as $n$ decreases further. For lower values of $n$ or larger $p$, the system becomes superconducting again, but now via the $d$-wave channel. The second peak in the order parameter is shown by $\Phi_d$ at $n\sim 0.8$. Thus a double-hump feature arises
due to overlap of the two domes at intermediate carrier concentration.
Very interestingly, a similar double-hump feature was also observed experimentally in the  variation of critical superconducting temperature $T_c$ as a function of doping
~\cite{Zeng2020PhaseDiagram,Chow2022InfiniteLayer,Li2020SuperconductingDome,Osada2022FieldEffect}. 
Our results show (cf. Fig.~\ref{fig:vmc-results}(a)) that the nature of superconductivity in the two domes are drastically different. 
For smaller hole doping, the system behaves as three-dimensional superconductor with extended
$s$-wave type pairing, whereas at higher hole doping it is of two-dimensional nature with $d$-wave 
pairing as in the cuprates. The origin of the double-dome structure in SC correlations can be
qualitatively understood by looking at the individual electron densities in the two bands. 
In Fig.~\ref{fig:vmc-results}(b), we plot $n_{\alpha}=\langle N_{\alpha}\rangle/L$ as a function
of $n$, where $N_{\alpha}$ is the number of electrons in orbital $\alpha$.
The figure shows that, as holes are introduced into the lattice, it first goes to the IS orbitals while the $d_{x^2-y^2}$ orbitals remain undoped. For example, at $n=1.6$ where $\Phi_s$ is peaked, we have
$n_s\sim 0.6$ whereas $n_d\sim 1.0$. Thus, the presence of superconductivity in the IS band and its
absence in the $d_{x^2-y^2}$ band in this regime of $n$ can be understood to be of the same origin 
as in hole doped cuprates. The optimal doping for the IS band ($\sim 0.4$) though turns out to be significantly larger than that in cuprates. 
For smaller value of $n$, e.g.~at $n=0.8$, $n_d\sim 0.7$ but $n_s\sim 0.1$. Thus, the $d_{x^2-y^2}$ band becomes optimally doped and $\Phi_d$ peaks. The IS band loses superconductivity due to loss of pairing as $n_s$ becomes very small. 
Moving to effect of H, as seen in Fig.~\ref{fig:vmc-results}(a),  
introduction of hydrogen impurity affects the two superconducting domes differently.  While the hydrogenation enhances the $s$-wave SC correlations in larger particle density regime, there is only a slight reduction in $d$-wave SC correlations at smaller values of $n$. This can be tentatively understood as follows. Ignoring inter-orbital hoppings, the effective superexchange 
coupling corresponding to hopping between nearest neighbour (NN) sites can be defined as $J_{\alpha} \sim \frac{t^2_{\alpha}}{U_{\alpha}}$, where $t_{\alpha}$ is the NN hopping amplitude
and $U_{\alpha}$ is the Hubbard interaction for orbital $\alpha$. 
From Fig.~\ref{fig:hopping-parameters}(b), 
we see that the magnitude of $t_{s}$ increases significantly while that of $t_{d}$ decreases
in the hydrogenated compound, compared to the dehydrogenated case. This results in a stronger $J_{s}$ and slightly weaker
$J_{d}$ in the H-bearing system. Since the pairing correlations are proportional to $J_{\alpha}$,
the SC correlations are also affected accordingly as found in our results.

\section{Summary and Discussion}
\label{sec:conclusion} 

In summary, motivated by the recent discussions~\cite{Ding2023HydrogenNickelates,Si2020} on the possible influence
of topotatic hydrogen in the superconductivity of infinite-layer nickelate, using
a combination of first-principles and variational Monte Carlo techniques, we theoretically investigate this issue. Our study has two important findings,
firstly, the minimalistic description of nickelate remains two-band even in
the presence of H. While H affects the local IS Wannier function for the Ni strongly bonded to H, it does not have a significant influence on $d_{x^2-y^2}$ Wannier function. The Fermi surface of the hydrogenated compound at $k_z$ = 0 plane 
shows similar topology as that of the dehydrogenated compound, while changes
are observed for that at $k_z$ = $\pi$ plane in terms of near disappearance of
electron pocket arising from IS. The electron density-dependence of the calculated superconducting order parameter shows similar features for H-bearing as well as pristine compound, that includes a two-hump nature arising out of overlap of two domes dominated by IS at low doping regime and by $d_{x^2-y^2}$ at high doping regime. The hydrogenation strengthens the superconductivity from IS, leaving
$d_{x^2-y^2}$ contribution more or less unchanged. This supports the recent
view~\cite{Balakrishnan2024HydrogenNickelates, Gonzalez2024HydrogenNickelates} that hydrogen incorporation is not essential for superconductivity in topotactically reduced nickelates.

Finally, experimentally, superconductivity is observed over a rather narrow region of doping. Our theoretical results, on the other hand, find the existence of superconductivity for a much broader doping range. While this may arise from several
sources of inaccuracies in our calculations, e.g. neglect of spin-exchange and pair-hopping terms in the Hamiltonian, choice of interaction parameters, the system size dependence,
our implementation also may miss details of the experimental set-up, like substrate effect and capping. 
Some studies~\cite{10.1063/5.0005103,Krieger2022Nickelate,Krieger_2023,10.1063/5.0197304} suggest that capping layers, often used to protect nickelate films during synthesis, can significantly influence their superconducting properties, though it remains debated. 
Further investigation is needed to resolve this.

\section*{Acknowledgement} 
The authors acknowledge computing facilities in the Baryon and Param Rudra clusters 
at the SNBNCBS HPC center. M.G. acknowledges CSIR, India, for the senior research fellowship (Grant no. 09/575 (0131) 2020-EMR-I).
T.S.D. acknowledges JC Bose National Fellowship (Grant No. JCB/2020/000004) for funding. The authors acknowledge discussions with Arijit Haldar and Tanmoy Das.

\clearpage        
\onecolumngrid
\appendix
\section{Effect of Hydrogen Impurity on Ni-\textit{d} Levels}
\label{ref:appnxa}

\begin{figure*}[htb]
    \centering
    \includegraphics[width=1\linewidth]{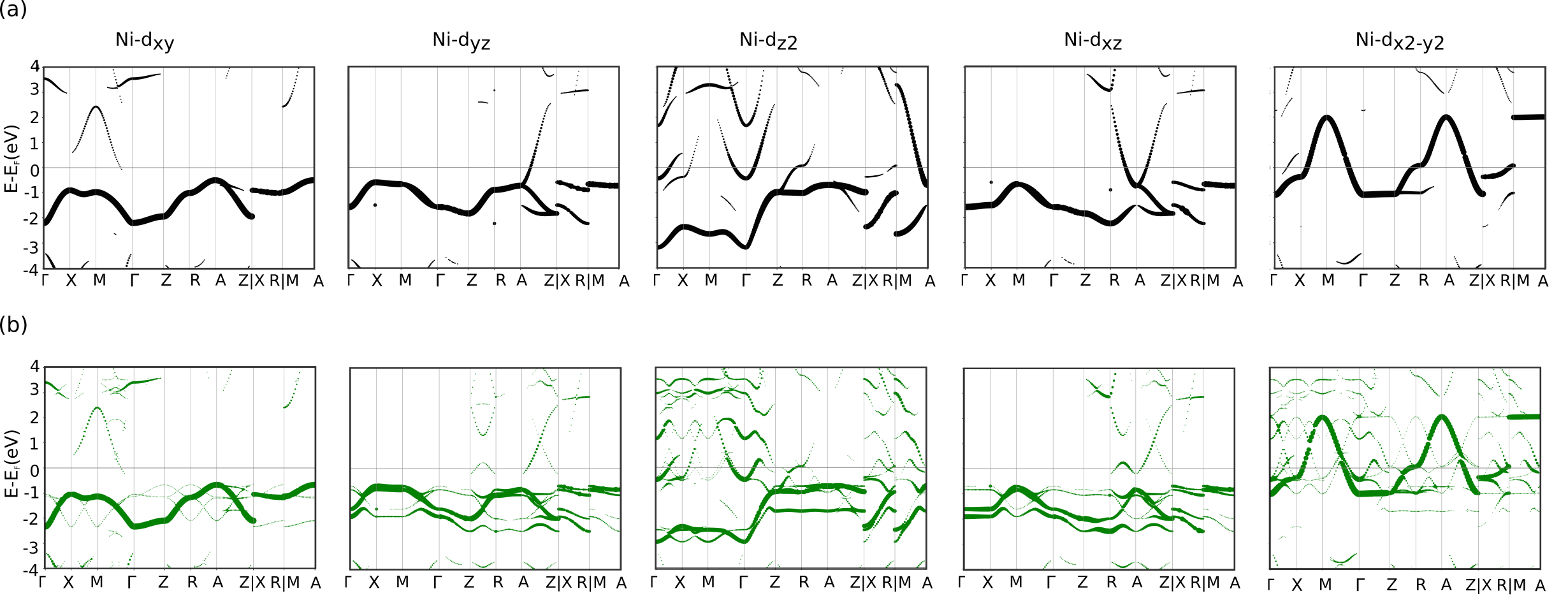}
    \caption{Orbital-projected band structure for the Ni-\textit{d} orbitals. (a) dehydrogenated case. (b) Ni$_{AOV}$ states for the hydrogenated case.}
    \label{fig:band_character}
\end{figure*}

\clearpage
\twocolumngrid
\bibliography{ref}  

\begin{thebibliography}{80}%
\makeatletter
\providecommand \@ifxundefined [1]{%
 \@ifx{#1\undefined}
}%
\providecommand \@ifnum [1]{%
 \ifnum #1\expandafter \@firstoftwo
 \else \expandafter \@secondoftwo
 \fi
}%
\providecommand \@ifx [1]{%
 \ifx #1\expandafter \@firstoftwo
 \else \expandafter \@secondoftwo
 \fi
}%
\providecommand \natexlab [1]{#1}%
\providecommand \enquote  [1]{``#1''}%
\providecommand \bibnamefont  [1]{#1}%
\providecommand \bibfnamefont [1]{#1}%
\providecommand \citenamefont [1]{#1}%
\providecommand \href@noop [0]{\@secondoftwo}%
\providecommand \href [0]{\begingroup \@sanitize@url \@href}%
\providecommand \@href[1]{\@@startlink{#1}\@@href}%
\providecommand \@@href[1]{\endgroup#1\@@endlink}%
\providecommand \@sanitize@url [0]{\catcode `\\12\catcode `\$12\catcode `\&12\catcode `\#12\catcode `\^12\catcode `\_12\catcode `\%12\relax}%
\providecommand \@@startlink[1]{}%
\providecommand \@@endlink[0]{}%
\providecommand \url  [0]{\begingroup\@sanitize@url \@url }%
\providecommand \@url [1]{\endgroup\@href {#1}{\urlprefix }}%
\providecommand \urlprefix  [0]{URL }%
\providecommand \Eprint [0]{\href }%
\providecommand \doibase [0]{https://doi.org/}%
\providecommand \selectlanguage [0]{\@gobble}%
\providecommand \bibinfo  [0]{\@secondoftwo}%
\providecommand \bibfield  [0]{\@secondoftwo}%
\providecommand \translation [1]{[#1]}%
\providecommand \BibitemOpen [0]{}%
\providecommand \bibitemStop [0]{}%
\providecommand \bibitemNoStop [0]{.\EOS\space}%
\providecommand \EOS [0]{\spacefactor3000\relax}%
\providecommand \BibitemShut  [1]{\csname bibitem#1\endcsname}%
\let\auto@bib@innerbib\@empty
\bibitem [{\citenamefont {Bednorz}\ and\ \citenamefont {Müller}(1986)}]{Bednorz1986}%
  \BibitemOpen
  \bibfield  {author} {\bibinfo {author} {\bibfnamefont {J.~G.}\ \bibnamefont {Bednorz}}\ and\ \bibinfo {author} {\bibfnamefont {K.~A.}\ \bibnamefont {Müller}},\ }\bibfield  {title} {\bibinfo {title} {Possible high{Tc} superconductivity in the {Ba-La-Cu-O} system},\ }\href {https://doi.org/10.1007/BF01303701} {\bibfield  {journal} {\bibinfo  {journal} {Zeitschrift für Physik B Condensed Matter}\ }\textbf {\bibinfo {volume} {64}},\ \bibinfo {pages} {189} (\bibinfo {year} {1986})}\BibitemShut {NoStop}%
\bibitem [{\citenamefont {Anderson}(1987)}]{anderson1987resonating}%
  \BibitemOpen
  \bibfield  {author} {\bibinfo {author} {\bibfnamefont {P.~W.}\ \bibnamefont {Anderson}},\ }\bibfield  {title} {\bibinfo {title} {The resonating valence bond state in {La$_2$CuO$_4$} and superconductivity},\ }\href {https://doi.org/10.1126/science.235.4793.1196} {\bibfield  {journal} {\bibinfo  {journal} {Science}\ }\textbf {\bibinfo {volume} {235}},\ \bibinfo {pages} {1196} (\bibinfo {year} {1987})}\BibitemShut {NoStop}%
\bibitem [{\citenamefont {Dagotto}(1994)}]{RevModPhys.66.763}%
  \BibitemOpen
  \bibfield  {author} {\bibinfo {author} {\bibfnamefont {E.}~\bibnamefont {Dagotto}},\ }\bibfield  {title} {\bibinfo {title} {Correlated electrons in high-temperature superconductors},\ }\href {https://doi.org/10.1103/RevModPhys.66.763} {\bibfield  {journal} {\bibinfo  {journal} {Rev. Mod. Phys.}\ }\textbf {\bibinfo {volume} {66}},\ \bibinfo {pages} {763} (\bibinfo {year} {1994})}\BibitemShut {NoStop}%
\bibitem [{\citenamefont {Tsuei}\ and\ \citenamefont {Kirtley}(2000)}]{RevModPhys.72.969}%
  \BibitemOpen
  \bibfield  {author} {\bibinfo {author} {\bibfnamefont {C.~C.}\ \bibnamefont {Tsuei}}\ and\ \bibinfo {author} {\bibfnamefont {J.~R.}\ \bibnamefont {Kirtley}},\ }\bibfield  {title} {\bibinfo {title} {Pairing symmetry in cuprate superconductors},\ }\href {https://doi.org/10.1103/RevModPhys.72.969} {\bibfield  {journal} {\bibinfo  {journal} {Rev. Mod. Phys.}\ }\textbf {\bibinfo {volume} {72}},\ \bibinfo {pages} {969} (\bibinfo {year} {2000})}\BibitemShut {NoStop}%
\bibitem [{\citenamefont {Taillefer}(2010)}]{annurevLouis}%
  \BibitemOpen
  \bibfield  {author} {\bibinfo {author} {\bibfnamefont {L.}~\bibnamefont {Taillefer}},\ }\bibfield  {title} {\bibinfo {title} {{Scattering and Pairing in Cuprate Superconductors}},\ }\href {https://doi.org/https://doi.org/10.1146/annurev-conmatphys-070909-104117} {\bibfield  {journal} {\bibinfo  {journal} {Annual Review of Condensed Matter Physics}\ }\textbf {\bibinfo {volume} {1}},\ \bibinfo {pages} {51} (\bibinfo {year} {2010})}\BibitemShut {NoStop}%
\bibitem [{\citenamefont {Siegrist}\ \emph {et~al.}(1988)\citenamefont {Siegrist}, \citenamefont {Zahurak}, \citenamefont {Murphy},\ and\ \citenamefont {Roth}}]{siegrist1988crystal}%
  \BibitemOpen
  \bibfield  {author} {\bibinfo {author} {\bibfnamefont {T.}~\bibnamefont {Siegrist}}, \bibinfo {author} {\bibfnamefont {S.~M.}\ \bibnamefont {Zahurak}}, \bibinfo {author} {\bibfnamefont {D.~W.}\ \bibnamefont {Murphy}},\ and\ \bibinfo {author} {\bibfnamefont {R.~S.}\ \bibnamefont {Roth}},\ }\bibfield  {title} {\bibinfo {title} {The crystal structure of the superconducting compound {La$_{2-x}$Ba$_x$CuO$_4$}},\ }\href {https://doi.org/10.1038/334231a0} {\bibfield  {journal} {\bibinfo  {journal} {Nature}\ }\textbf {\bibinfo {volume} {334}},\ \bibinfo {pages} {231} (\bibinfo {year} {1988})}\BibitemShut {NoStop}%
\bibitem [{\citenamefont {Balestrino}\ \emph {et~al.}(2002)\citenamefont {Balestrino}, \citenamefont {Medaglia}, \citenamefont {Orgiani}, \citenamefont {Tebano}, \citenamefont {Aruta}, \citenamefont {Lavanga},\ and\ \citenamefont {Varlamov}}]{balestrino2002evidence}%
  \BibitemOpen
  \bibfield  {author} {\bibinfo {author} {\bibfnamefont {G.}~\bibnamefont {Balestrino}}, \bibinfo {author} {\bibfnamefont {P.~G.}\ \bibnamefont {Medaglia}}, \bibinfo {author} {\bibfnamefont {P.}~\bibnamefont {Orgiani}}, \bibinfo {author} {\bibfnamefont {A.}~\bibnamefont {Tebano}}, \bibinfo {author} {\bibfnamefont {C.}~\bibnamefont {Aruta}}, \bibinfo {author} {\bibfnamefont {S.}~\bibnamefont {Lavanga}},\ and\ \bibinfo {author} {\bibfnamefont {A.~A.}\ \bibnamefont {Varlamov}},\ }\bibfield  {title} {\bibinfo {title} {{Evidence of Two-Dimensional Nernst Signal in Epitaxial Cuprate Thin Films}},\ }\href {https://doi.org/10.1103/PhysRevLett.89.156402} {\bibfield  {journal} {\bibinfo  {journal} {Physical Review Letters}\ }\textbf {\bibinfo {volume} {89}},\ \bibinfo {pages} {156402} (\bibinfo {year} {2002})}\BibitemShut {NoStop}%
\bibitem [{\citenamefont {Di~Castro}\ \emph {et~al.}(2015)\citenamefont {Di~Castro}, \citenamefont {Cantoni}, \citenamefont {Ridolfi}, \citenamefont {Aruta}, \citenamefont {Tebano}, \citenamefont {Yang},\ and\ \citenamefont {Balestrino}}]{dicastro2015direct}%
  \BibitemOpen
  \bibfield  {author} {\bibinfo {author} {\bibfnamefont {D.}~\bibnamefont {Di~Castro}}, \bibinfo {author} {\bibfnamefont {C.}~\bibnamefont {Cantoni}}, \bibinfo {author} {\bibfnamefont {F.}~\bibnamefont {Ridolfi}}, \bibinfo {author} {\bibfnamefont {C.}~\bibnamefont {Aruta}}, \bibinfo {author} {\bibfnamefont {A.}~\bibnamefont {Tebano}}, \bibinfo {author} {\bibfnamefont {N.}~\bibnamefont {Yang}},\ and\ \bibinfo {author} {\bibfnamefont {G.}~\bibnamefont {Balestrino}},\ }\bibfield  {title} {\bibinfo {title} {{Direct Observation of Charge Density Wave in Cuprate Superconductors}},\ }\href {https://doi.org/10.1103/PhysRevLett.115.147001} {\bibfield  {journal} {\bibinfo  {journal} {Physical Review Letters}\ }\textbf {\bibinfo {volume} {115}},\ \bibinfo {pages} {147001} (\bibinfo {year} {2015})}\BibitemShut {NoStop}%
\bibitem [{\citenamefont {Kamihara}\ \emph {et~al.}(2006)\citenamefont {Kamihara}, \citenamefont {Hiramatsu}, \citenamefont {Hirano}, \citenamefont {Kawamura}, \citenamefont {Yanagi}, \citenamefont {Kamiya},\ and\ \citenamefont {Hosono}}]{doi:10.1021/ja063355c}%
  \BibitemOpen
  \bibfield  {author} {\bibinfo {author} {\bibfnamefont {Y.}~\bibnamefont {Kamihara}}, \bibinfo {author} {\bibfnamefont {H.}~\bibnamefont {Hiramatsu}}, \bibinfo {author} {\bibfnamefont {M.}~\bibnamefont {Hirano}}, \bibinfo {author} {\bibfnamefont {R.}~\bibnamefont {Kawamura}}, \bibinfo {author} {\bibfnamefont {H.}~\bibnamefont {Yanagi}}, \bibinfo {author} {\bibfnamefont {T.}~\bibnamefont {Kamiya}},\ and\ \bibinfo {author} {\bibfnamefont {H.}~\bibnamefont {Hosono}},\ }\bibfield  {title} {\bibinfo {title} {{Iron-Based Layered Superconductor: LaOFeP}},\ }\href {https://doi.org/10.1021/ja063355c} {\bibfield  {journal} {\bibinfo  {journal} {Journal of the American Chemical Society}\ }\textbf {\bibinfo {volume} {128}},\ \bibinfo {pages} {10012} (\bibinfo {year} {2006})}\BibitemShut {NoStop}%
\bibitem [{\citenamefont {Kamihara}\ \emph {et~al.}(2008)\citenamefont {Kamihara}, \citenamefont {Watanabe}, \citenamefont {Hirano},\ and\ \citenamefont {Hosono}}]{doi:10.1021/ja800073m}%
  \BibitemOpen
  \bibfield  {author} {\bibinfo {author} {\bibfnamefont {Y.}~\bibnamefont {Kamihara}}, \bibinfo {author} {\bibfnamefont {T.}~\bibnamefont {Watanabe}}, \bibinfo {author} {\bibfnamefont {M.}~\bibnamefont {Hirano}},\ and\ \bibinfo {author} {\bibfnamefont {H.}~\bibnamefont {Hosono}},\ }\bibfield  {title} {\bibinfo {title} {{Iron-Based Layered Superconductor {La[O$_{1-x}$F$_x$]FeAs} (x = 0.05-0.12) with {T$_c$} = 26 K}},\ }\href {https://doi.org/10.1021/ja800073m} {\bibfield  {journal} {\bibinfo  {journal} {Journal of the American Chemical Society}\ }\textbf {\bibinfo {volume} {130}},\ \bibinfo {pages} {3296} (\bibinfo {year} {2008})}\BibitemShut {NoStop}%
\bibitem [{\citenamefont {Mizuguchi}\ and\ \citenamefont {Takano}(2010)}]{doi:10.1143/JPSJ.79.102001}%
  \BibitemOpen
  \bibfield  {author} {\bibinfo {author} {\bibfnamefont {Y.}~\bibnamefont {Mizuguchi}}\ and\ \bibinfo {author} {\bibfnamefont {Y.}~\bibnamefont {Takano}},\ }\bibfield  {title} {\bibinfo {title} {{Review of {Fe} Chalcogenides as the Simplest {Fe}-Based Superconductor}},\ }\href {https://doi.org/10.1143/JPSJ.79.102001} {\bibfield  {journal} {\bibinfo  {journal} {Journal of the Physical Society of Japan}\ }\textbf {\bibinfo {volume} {79}},\ \bibinfo {pages} {102001} (\bibinfo {year} {2010})}\BibitemShut {NoStop}%
\bibitem [{\citenamefont {Stewart}(2011)}]{RevModPhys.83.1589}%
  \BibitemOpen
  \bibfield  {author} {\bibinfo {author} {\bibfnamefont {G.~R.}\ \bibnamefont {Stewart}},\ }\bibfield  {title} {\bibinfo {title} {Superconductivity in iron compounds},\ }\href {https://doi.org/10.1103/RevModPhys.83.1589} {\bibfield  {journal} {\bibinfo  {journal} {Rev. Mod. Phys.}\ }\textbf {\bibinfo {volume} {83}},\ \bibinfo {pages} {1589} (\bibinfo {year} {2011})}\BibitemShut {NoStop}%
\bibitem [{\citenamefont {Chaloupka}\ and\ \citenamefont {Khaliullin}(2008)}]{chaloupka2008orbital}%
  \BibitemOpen
  \bibfield  {author} {\bibinfo {author} {\bibfnamefont {J.}~\bibnamefont {Chaloupka}}\ and\ \bibinfo {author} {\bibfnamefont {G.}~\bibnamefont {Khaliullin}},\ }\bibfield  {title} {\bibinfo {title} {Orbital order and possible superconductivity in {LaNiO$_3$/LaMO$_3$} superlattices},\ }\href {https://doi.org/10.1103/PhysRevLett.100.016404} {\bibfield  {journal} {\bibinfo  {journal} {Physical Review Letters}\ }\textbf {\bibinfo {volume} {100}},\ \bibinfo {pages} {016404} (\bibinfo {year} {2008})}\BibitemShut {NoStop}%
\bibitem [{\citenamefont {Hansmann}\ \emph {et~al.}(2009)\citenamefont {Hansmann}, \citenamefont {Yang}, \citenamefont {Toschi}, \citenamefont {Khaliullin}, \citenamefont {Andersen},\ and\ \citenamefont {Held}}]{hansmann2009turning}%
  \BibitemOpen
  \bibfield  {author} {\bibinfo {author} {\bibfnamefont {P.}~\bibnamefont {Hansmann}}, \bibinfo {author} {\bibfnamefont {X.}~\bibnamefont {Yang}}, \bibinfo {author} {\bibfnamefont {A.}~\bibnamefont {Toschi}}, \bibinfo {author} {\bibfnamefont {G.}~\bibnamefont {Khaliullin}}, \bibinfo {author} {\bibfnamefont {O.~K.}\ \bibnamefont {Andersen}},\ and\ \bibinfo {author} {\bibfnamefont {K.}~\bibnamefont {Held}},\ }\bibfield  {title} {\bibinfo {title} {{Turning a nickelate Fermi surface into a cupratelike one through heterostructuring}},\ }\href {https://doi.org/10.1103/PhysRevLett.103.016401} {\bibfield  {journal} {\bibinfo  {journal} {Physical Review Letters}\ }\textbf {\bibinfo {volume} {103}},\ \bibinfo {pages} {016401} (\bibinfo {year} {2009})}\BibitemShut {NoStop}%
\bibitem [{\citenamefont {Poltavets}\ \emph {et~al.}(2006)\citenamefont {Poltavets}, \citenamefont {Lokshin}, \citenamefont {Dikmen}, \citenamefont {Croft}, \citenamefont {Egami},\ and\ \citenamefont {Greenblatt}}]{doi:10.1021/ja063031o}%
  \BibitemOpen
  \bibfield  {author} {\bibinfo {author} {\bibfnamefont {V.~V.}\ \bibnamefont {Poltavets}}, \bibinfo {author} {\bibfnamefont {K.~A.}\ \bibnamefont {Lokshin}}, \bibinfo {author} {\bibfnamefont {S.}~\bibnamefont {Dikmen}}, \bibinfo {author} {\bibfnamefont {M.}~\bibnamefont {Croft}}, \bibinfo {author} {\bibfnamefont {T.}~\bibnamefont {Egami}},\ and\ \bibinfo {author} {\bibfnamefont {M.}~\bibnamefont {Greenblatt}},\ }\bibfield  {title} {\bibinfo {title} {{La$_3$Ni$_2$O$_6$: A New Double T'-type Nickelate with Infinite Ni$^{1+/2+}$O$_2$ Layers}},\ }\href {https://doi.org/10.1021/ja063031o} {\bibfield  {journal} {\bibinfo  {journal} {Journal of the American Chemical Society}\ }\textbf {\bibinfo {volume} {128}},\ \bibinfo {pages} {9050} (\bibinfo {year} {2006})}\BibitemShut {NoStop}%
\bibitem [{\citenamefont {Poltavets}\ \emph {et~al.}(2007)\citenamefont {Poltavets}, \citenamefont {Lokshin}, \citenamefont {Croft}, \citenamefont {Mandal}, \citenamefont {Egami},\ and\ \citenamefont {Greenblatt}}]{doi:10.1021/ic701480v}%
  \BibitemOpen
  \bibfield  {author} {\bibinfo {author} {\bibfnamefont {V.~V.}\ \bibnamefont {Poltavets}}, \bibinfo {author} {\bibfnamefont {K.~A.}\ \bibnamefont {Lokshin}}, \bibinfo {author} {\bibfnamefont {M.}~\bibnamefont {Croft}}, \bibinfo {author} {\bibfnamefont {T.~K.}\ \bibnamefont {Mandal}}, \bibinfo {author} {\bibfnamefont {T.}~\bibnamefont {Egami}},\ and\ \bibinfo {author} {\bibfnamefont {M.}~\bibnamefont {Greenblatt}},\ }\bibfield  {title} {\bibinfo {title} {{Crystal Structures of Ln$_4$Ni$_3$O$_8$ (Ln = La, Nd) Triple Layer T'-type Nickelates}},\ }\href {https://doi.org/10.1021/ic701480v} {\bibfield  {journal} {\bibinfo  {journal} {Inorganic Chemistry}\ }\textbf {\bibinfo {volume} {46}},\ \bibinfo {pages} {10887} (\bibinfo {year} {2007})}\BibitemShut {NoStop}%
\bibitem [{\citenamefont {Sarkar}\ \emph {et~al.}(2011)\citenamefont {Sarkar}, \citenamefont {Dasgupta}, \citenamefont {Greenblatt},\ and\ \citenamefont {Saha-Dasgupta}}]{PhysRevB.84.180411}%
  \BibitemOpen
  \bibfield  {author} {\bibinfo {author} {\bibfnamefont {S.}~\bibnamefont {Sarkar}}, \bibinfo {author} {\bibfnamefont {I.}~\bibnamefont {Dasgupta}}, \bibinfo {author} {\bibfnamefont {M.}~\bibnamefont {Greenblatt}},\ and\ \bibinfo {author} {\bibfnamefont {T.}~\bibnamefont {Saha-Dasgupta}},\ }\bibfield  {title} {\bibinfo {title} {{Electronic and magnetic structures of bilayer La${}_{3}$Ni${}_{2}$O${}_{6}$ and trilayer La${}_{4}$Ni${}_{3}$O${}_{8}$ nickelates from first principles}},\ }\href {https://doi.org/10.1103/PhysRevB.84.180411} {\bibfield  {journal} {\bibinfo  {journal} {Phys. Rev. B}\ }\textbf {\bibinfo {volume} {84}},\ \bibinfo {pages} {180411} (\bibinfo {year} {2011})}\BibitemShut {NoStop}%
\bibitem [{\citenamefont {Li}\ \emph {et~al.}(2019)\citenamefont {Li}, \citenamefont {Lee}, \citenamefont {Wang}, \citenamefont {Osada}, \citenamefont {Crossley}, \citenamefont {Lee}, \citenamefont {Cui}, \citenamefont {Hikita},\ and\ \citenamefont {Hwang}}]{nature2019nickelate}%
  \BibitemOpen
  \bibfield  {author} {\bibinfo {author} {\bibfnamefont {D.}~\bibnamefont {Li}}, \bibinfo {author} {\bibfnamefont {K.}~\bibnamefont {Lee}}, \bibinfo {author} {\bibfnamefont {B.~Y.}\ \bibnamefont {Wang}}, \bibinfo {author} {\bibfnamefont {M.}~\bibnamefont {Osada}}, \bibinfo {author} {\bibfnamefont {S.}~\bibnamefont {Crossley}}, \bibinfo {author} {\bibfnamefont {H.~R.}\ \bibnamefont {Lee}}, \bibinfo {author} {\bibfnamefont {Y.}~\bibnamefont {Cui}}, \bibinfo {author} {\bibfnamefont {Y.}~\bibnamefont {Hikita}},\ and\ \bibinfo {author} {\bibfnamefont {H.~Y.}\ \bibnamefont {Hwang}},\ }\bibfield  {title} {\bibinfo {title} {Superconductivity in an infinite-layer nickelate},\ }\href {https://doi.org/10.1038/s41586-019-1496-5} {\bibfield  {journal} {\bibinfo  {journal} {Nature}\ }\textbf {\bibinfo {volume} {572}},\ \bibinfo {pages} {624} (\bibinfo {year} {2019})}\BibitemShut {NoStop}%
\bibitem [{\citenamefont {Nomura}\ and\ \citenamefont {Arita}(2022)}]{nomura2022superconductivity}%
  \BibitemOpen
  \bibfield  {author} {\bibinfo {author} {\bibfnamefont {Y.}~\bibnamefont {Nomura}}\ and\ \bibinfo {author} {\bibfnamefont {R.}~\bibnamefont {Arita}},\ }\bibfield  {title} {\bibinfo {title} {Superconductivity in infinite-layer nickelates},\ }\href {https://doi.org/10.1088/1361-6633/ac5a60} {\bibfield  {journal} {\bibinfo  {journal} {Reports on Progress in Physics}\ }\textbf {\bibinfo {volume} {85}},\ \bibinfo {pages} {052501} (\bibinfo {year} {2022})}\BibitemShut {NoStop}%
\bibitem [{\citenamefont {Osada}\ \emph {et~al.}(2021)\citenamefont {Osada}, \citenamefont {Wang}, \citenamefont {Goodge}, \citenamefont {Harvey}, \citenamefont {Lee}, \citenamefont {Li}, \citenamefont {Kourkoutis},\ and\ \citenamefont {Hwang}}]{osada2021nickelate}%
  \BibitemOpen
  \bibfield  {author} {\bibinfo {author} {\bibfnamefont {M.}~\bibnamefont {Osada}}, \bibinfo {author} {\bibfnamefont {B.~Y.}\ \bibnamefont {Wang}}, \bibinfo {author} {\bibfnamefont {B.~H.}\ \bibnamefont {Goodge}}, \bibinfo {author} {\bibfnamefont {S.~P.}\ \bibnamefont {Harvey}}, \bibinfo {author} {\bibfnamefont {K.}~\bibnamefont {Lee}}, \bibinfo {author} {\bibfnamefont {D.}~\bibnamefont {Li}}, \bibinfo {author} {\bibfnamefont {L.~F.}\ \bibnamefont {Kourkoutis}},\ and\ \bibinfo {author} {\bibfnamefont {H.~Y.}\ \bibnamefont {Hwang}},\ }\bibfield  {title} {\bibinfo {title} {{Nickelate Superconductivity without Rare-Earth Magnetism: (La,Sr)NiO$_2$}},\ }\href {https://doi.org/10.1002/adma.202104083} {\bibfield  {journal} {\bibinfo  {journal} {Advanced Materials}\ }\textbf {\bibinfo {volume} {33}},\ \bibinfo {pages} {2104083} (\bibinfo {year} {2021})}\BibitemShut {NoStop}%
\bibitem [{\citenamefont {Norman}(2020)}]{Norman2020NickelAge}%
  \BibitemOpen
  \bibfield  {author} {\bibinfo {author} {\bibfnamefont {M.~R.}\ \bibnamefont {Norman}},\ }\bibfield  {title} {\bibinfo {title} {{Entering the Nickel Age of Superconductivity}},\ }\href {https://physics.aps.org/articles/v13/85} {\bibfield  {journal} {\bibinfo  {journal} {Physics}\ }\textbf {\bibinfo {volume} {13}},\ \bibinfo {pages} {85} (\bibinfo {year} {2020})}\BibitemShut {NoStop}%
\bibitem [{\citenamefont {Lee}\ \emph {et~al.}(2006)\citenamefont {Lee}, \citenamefont {Nagaosa},\ and\ \citenamefont {Wen}}]{lee2006doping}%
  \BibitemOpen
  \bibfield  {author} {\bibinfo {author} {\bibfnamefont {P.~A.}\ \bibnamefont {Lee}}, \bibinfo {author} {\bibfnamefont {N.}~\bibnamefont {Nagaosa}},\ and\ \bibinfo {author} {\bibfnamefont {X.-G.}\ \bibnamefont {Wen}},\ }\bibfield  {title} {\bibinfo {title} {{Doping a Mott insulator: Physics of high-temperature superconductivity}},\ }\href {https://doi.org/10.1103/RevModPhys.78.17} {\bibfield  {journal} {\bibinfo  {journal} {Reviews of Modern Physics}\ }\textbf {\bibinfo {volume} {78}},\ \bibinfo {pages} {17} (\bibinfo {year} {2006})}\BibitemShut {NoStop}%
\bibitem [{\citenamefont {Hsu}\ \emph {et~al.}(2021)\citenamefont {Hsu}, \citenamefont {Wang}, \citenamefont {Berben}, \citenamefont {Li}, \citenamefont {Lee}, \citenamefont {Duffy}, \citenamefont {Ottenbros}, \citenamefont {Kim}, \citenamefont {Osada}, \citenamefont {Wiedmann}, \citenamefont {Hwang},\ and\ \citenamefont {Hussey}}]{PhysRevResearch.3.L042015}%
  \BibitemOpen
  \bibfield  {author} {\bibinfo {author} {\bibfnamefont {Y.-T.}\ \bibnamefont {Hsu}}, \bibinfo {author} {\bibfnamefont {B.~Y.}\ \bibnamefont {Wang}}, \bibinfo {author} {\bibfnamefont {M.}~\bibnamefont {Berben}}, \bibinfo {author} {\bibfnamefont {D.}~\bibnamefont {Li}}, \bibinfo {author} {\bibfnamefont {K.}~\bibnamefont {Lee}}, \bibinfo {author} {\bibfnamefont {C.}~\bibnamefont {Duffy}}, \bibinfo {author} {\bibfnamefont {T.}~\bibnamefont {Ottenbros}}, \bibinfo {author} {\bibfnamefont {W.~J.}\ \bibnamefont {Kim}}, \bibinfo {author} {\bibfnamefont {M.}~\bibnamefont {Osada}}, \bibinfo {author} {\bibfnamefont {S.}~\bibnamefont {Wiedmann}}, \bibinfo {author} {\bibfnamefont {H.~Y.}\ \bibnamefont {Hwang}},\ and\ \bibinfo {author} {\bibfnamefont {N.~E.}\ \bibnamefont {Hussey}},\ }\bibfield  {title} {\bibinfo {title} {Insulator-to-metal crossover near the edge of the superconducting dome in {${\mathrm{Nd}}_{1\ensuremath{-}x}{\mathrm{Sr}}_{x}{\mathrm{NiO}}_{2}$}},\ }\href
  {https://doi.org/10.1103/PhysRevResearch.3.L042015} {\bibfield  {journal} {\bibinfo  {journal} {Phys. Rev. Res.}\ }\textbf {\bibinfo {volume} {3}},\ \bibinfo {pages} {L042015} (\bibinfo {year} {2021})}\BibitemShut {NoStop}%
\bibitem [{\citenamefont {Rossi}\ \emph {et~al.}(2021)\citenamefont {Rossi}, \citenamefont {Lu}, \citenamefont {Nag}, \citenamefont {Li}, \citenamefont {Osada}, \citenamefont {Lee}, \citenamefont {Wang}, \citenamefont {Agrestini}, \citenamefont {Garcia-Fernandez}, \citenamefont {Kas}, \citenamefont {Chuang}, \citenamefont {Shen}, \citenamefont {Hwang}, \citenamefont {Moritz}, \citenamefont {Zhou}, \citenamefont {Devereaux},\ and\ \citenamefont {Lee}}]{PhysRevB.104.L220505}%
  \BibitemOpen
  \bibfield  {author} {\bibinfo {author} {\bibfnamefont {M.}~\bibnamefont {Rossi}}, \bibinfo {author} {\bibfnamefont {H.}~\bibnamefont {Lu}}, \bibinfo {author} {\bibfnamefont {A.}~\bibnamefont {Nag}}, \bibinfo {author} {\bibfnamefont {D.}~\bibnamefont {Li}}, \bibinfo {author} {\bibfnamefont {M.}~\bibnamefont {Osada}}, \bibinfo {author} {\bibfnamefont {K.}~\bibnamefont {Lee}}, \bibinfo {author} {\bibfnamefont {B.~Y.}\ \bibnamefont {Wang}}, \bibinfo {author} {\bibfnamefont {S.}~\bibnamefont {Agrestini}}, \bibinfo {author} {\bibfnamefont {M.}~\bibnamefont {Garcia-Fernandez}}, \bibinfo {author} {\bibfnamefont {J.~J.}\ \bibnamefont {Kas}}, \bibinfo {author} {\bibfnamefont {Y.-D.}\ \bibnamefont {Chuang}}, \bibinfo {author} {\bibfnamefont {Z.~X.}\ \bibnamefont {Shen}}, \bibinfo {author} {\bibfnamefont {H.~Y.}\ \bibnamefont {Hwang}}, \bibinfo {author} {\bibfnamefont {B.}~\bibnamefont {Moritz}}, \bibinfo {author} {\bibfnamefont {K.-J.}\ \bibnamefont {Zhou}}, \bibinfo {author} {\bibfnamefont {T.~P.}\ \bibnamefont
  {Devereaux}},\ and\ \bibinfo {author} {\bibfnamefont {W.~S.}\ \bibnamefont {Lee}},\ }\bibfield  {title} {\bibinfo {title} {Orbital and spin character of doped carriers in infinite-layer nickelates},\ }\href {https://doi.org/10.1103/PhysRevB.104.L220505} {\bibfield  {journal} {\bibinfo  {journal} {Phys. Rev. B}\ }\textbf {\bibinfo {volume} {104}},\ \bibinfo {pages} {L220505} (\bibinfo {year} {2021})}\BibitemShut {NoStop}%
\bibitem [{\citenamefont {Nomura}\ \emph {et~al.}(2019)\citenamefont {Nomura}, \citenamefont {Hirayama}, \citenamefont {Tadano}, \citenamefont {Yoshimoto}, \citenamefont {Nakamura},\ and\ \citenamefont {Arita}}]{nomura2019formation}%
  \BibitemOpen
  \bibfield  {author} {\bibinfo {author} {\bibfnamefont {Y.}~\bibnamefont {Nomura}}, \bibinfo {author} {\bibfnamefont {M.}~\bibnamefont {Hirayama}}, \bibinfo {author} {\bibfnamefont {T.}~\bibnamefont {Tadano}}, \bibinfo {author} {\bibfnamefont {Y.}~\bibnamefont {Yoshimoto}}, \bibinfo {author} {\bibfnamefont {K.}~\bibnamefont {Nakamura}},\ and\ \bibinfo {author} {\bibfnamefont {R.}~\bibnamefont {Arita}},\ }\bibfield  {title} {\bibinfo {title} {Formation of a two-dimensional single-component correlated electron system and band engineering in the nickelate superconductor {NdNiO$_2$}},\ }\href {https://doi.org/10.1103/PhysRevB.100.205138} {\bibfield  {journal} {\bibinfo  {journal} {Physical Review B}\ }\textbf {\bibinfo {volume} {100}},\ \bibinfo {pages} {205138} (\bibinfo {year} {2019})}\BibitemShut {NoStop}%
\bibitem [{\citenamefont {Lee}\ and\ \citenamefont {Pickett}(2004)}]{lee2004infinite}%
  \BibitemOpen
  \bibfield  {author} {\bibinfo {author} {\bibfnamefont {K.-W.}\ \bibnamefont {Lee}}\ and\ \bibinfo {author} {\bibfnamefont {W.~E.}\ \bibnamefont {Pickett}},\ }\bibfield  {title} {\bibinfo {title} {Infinite-layer {LaNiO$_2$: Ni$^{1+}$} is not {Cu$^{2+}$}},\ }\href {https://doi.org/10.1103/PhysRevB.70.165109} {\bibfield  {journal} {\bibinfo  {journal} {Physical Review B}\ }\textbf {\bibinfo {volume} {70}},\ \bibinfo {pages} {165109} (\bibinfo {year} {2004})}\BibitemShut {NoStop}%
\bibitem [{\citenamefont {Botana}\ and\ \citenamefont {Norman}(2020)}]{botana2020similarities}%
  \BibitemOpen
  \bibfield  {author} {\bibinfo {author} {\bibfnamefont {A.~S.}\ \bibnamefont {Botana}}\ and\ \bibinfo {author} {\bibfnamefont {M.~R.}\ \bibnamefont {Norman}},\ }\bibfield  {title} {\bibinfo {title} {{Similarities and Differences between {LaNiO$_2$} and {CaCuO$_2$} and Implications for Superconductivity}},\ }\href {https://doi.org/10.1103/PhysRevX.10.011024} {\bibfield  {journal} {\bibinfo  {journal} {Physical Review X}\ }\textbf {\bibinfo {volume} {10}},\ \bibinfo {pages} {011024} (\bibinfo {year} {2020})}\BibitemShut {NoStop}%
\bibitem [{\citenamefont {Sakakibara}\ \emph {et~al.}(2020)\citenamefont {Sakakibara}, \citenamefont {Usui}, \citenamefont {Suzuki}, \citenamefont {Takao}, \citenamefont {Aoki},\ and\ \citenamefont {Kuroki}}]{sakakibara2020model}%
  \BibitemOpen
  \bibfield  {author} {\bibinfo {author} {\bibfnamefont {H.}~\bibnamefont {Sakakibara}}, \bibinfo {author} {\bibfnamefont {H.}~\bibnamefont {Usui}}, \bibinfo {author} {\bibfnamefont {K.}~\bibnamefont {Suzuki}}, \bibinfo {author} {\bibfnamefont {K.}~\bibnamefont {Takao}}, \bibinfo {author} {\bibfnamefont {H.}~\bibnamefont {Aoki}},\ and\ \bibinfo {author} {\bibfnamefont {K.}~\bibnamefont {Kuroki}},\ }\bibfield  {title} {\bibinfo {title} {{Model Construction and a Possibility of Cupratelike Pairing in a New $d^9$ Nickelate Superconductor (Nd,Sr)NiO$_2$}},\ }\href {https://doi.org/10.1103/PhysRevLett.125.077003} {\bibfield  {journal} {\bibinfo  {journal} {Physical Review Letters}\ }\textbf {\bibinfo {volume} {125}},\ \bibinfo {pages} {077003} (\bibinfo {year} {2020})}\BibitemShut {NoStop}%
\bibitem [{\citenamefont {Hepting}\ \emph {et~al.}(2020)\citenamefont {Hepting}, \citenamefont {Li}, \citenamefont {Jia}, \citenamefont {Lu}, \citenamefont {Paris}, \citenamefont {Tseng}, \citenamefont {Feng}, \citenamefont {Osada}, \citenamefont {Been}, \citenamefont {Hikita}, \citenamefont {Chuang}, \citenamefont {Hussain}, \citenamefont {Zhou}, \citenamefont {Nag}, \citenamefont {Garcia-Fernandez}, \citenamefont {Rossi}, \citenamefont {Huang}, \citenamefont {Huang}, \citenamefont {Shen}, \citenamefont {Schmitt}, \citenamefont {Hwang}, \citenamefont {Moritz}, \citenamefont {Zaanen}, \citenamefont {Devereaux},\ and\ \citenamefont {Lee}}]{Hepting2020}%
  \BibitemOpen
  \bibfield  {author} {\bibinfo {author} {\bibfnamefont {M.}~\bibnamefont {Hepting}}, \bibinfo {author} {\bibfnamefont {D.}~\bibnamefont {Li}}, \bibinfo {author} {\bibfnamefont {C.~J.}\ \bibnamefont {Jia}}, \bibinfo {author} {\bibfnamefont {H.}~\bibnamefont {Lu}}, \bibinfo {author} {\bibfnamefont {E.}~\bibnamefont {Paris}}, \bibinfo {author} {\bibfnamefont {Y.}~\bibnamefont {Tseng}}, \bibinfo {author} {\bibfnamefont {X.}~\bibnamefont {Feng}}, \bibinfo {author} {\bibfnamefont {M.}~\bibnamefont {Osada}}, \bibinfo {author} {\bibfnamefont {E.}~\bibnamefont {Been}}, \bibinfo {author} {\bibfnamefont {Y.}~\bibnamefont {Hikita}}, \bibinfo {author} {\bibfnamefont {Y.-D.}\ \bibnamefont {Chuang}}, \bibinfo {author} {\bibfnamefont {Z.}~\bibnamefont {Hussain}}, \bibinfo {author} {\bibfnamefont {K.~J.}\ \bibnamefont {Zhou}}, \bibinfo {author} {\bibfnamefont {A.}~\bibnamefont {Nag}}, \bibinfo {author} {\bibfnamefont {M.}~\bibnamefont {Garcia-Fernandez}}, \bibinfo {author} {\bibfnamefont {M.}~\bibnamefont {Rossi}}, \bibinfo
  {author} {\bibfnamefont {H.~Y.}\ \bibnamefont {Huang}}, \bibinfo {author} {\bibfnamefont {D.~J.}\ \bibnamefont {Huang}}, \bibinfo {author} {\bibfnamefont {Z.~X.}\ \bibnamefont {Shen}}, \bibinfo {author} {\bibfnamefont {T.}~\bibnamefont {Schmitt}}, \bibinfo {author} {\bibfnamefont {H.~Y.}\ \bibnamefont {Hwang}}, \bibinfo {author} {\bibfnamefont {B.}~\bibnamefont {Moritz}}, \bibinfo {author} {\bibfnamefont {J.}~\bibnamefont {Zaanen}}, \bibinfo {author} {\bibfnamefont {T.~P.}\ \bibnamefont {Devereaux}},\ and\ \bibinfo {author} {\bibfnamefont {W.~S.}\ \bibnamefont {Lee}},\ }\bibfield  {title} {\bibinfo {title} {Electronic structure of the parent compound of superconducting infinite-layer nickelates},\ }\href {https://doi.org/10.1038/s41563-019-0585-z} {\bibfield  {journal} {\bibinfo  {journal} {Nature Materials}\ }\textbf {\bibinfo {volume} {19}},\ \bibinfo {pages} {381} (\bibinfo {year} {2020})}\BibitemShut {NoStop}%
\bibitem [{\citenamefont {Adhikary}\ \emph {et~al.}(2020)\citenamefont {Adhikary}, \citenamefont {Bandyopadhyay}, \citenamefont {Das}, \citenamefont {Dasgupta},\ and\ \citenamefont {Saha-Dasgupta}}]{PhysRevB.102.100501}%
  \BibitemOpen
  \bibfield  {author} {\bibinfo {author} {\bibfnamefont {P.}~\bibnamefont {Adhikary}}, \bibinfo {author} {\bibfnamefont {S.}~\bibnamefont {Bandyopadhyay}}, \bibinfo {author} {\bibfnamefont {T.}~\bibnamefont {Das}}, \bibinfo {author} {\bibfnamefont {I.}~\bibnamefont {Dasgupta}},\ and\ \bibinfo {author} {\bibfnamefont {T.}~\bibnamefont {Saha-Dasgupta}},\ }\bibfield  {title} {\bibinfo {title} {Orbital-selective superconductivity in a two-band model of infinite-layer nickelates},\ }\href {https://doi.org/10.1103/PhysRevB.102.100501} {\bibfield  {journal} {\bibinfo  {journal} {Phys. Rev. B}\ }\textbf {\bibinfo {volume} {102}},\ \bibinfo {pages} {100501} (\bibinfo {year} {2020})}\BibitemShut {NoStop}%
\bibitem [{\citenamefont {Bandyopadhyay}\ \emph {et~al.}(2020)\citenamefont {Bandyopadhyay}, \citenamefont {Adhikary}, \citenamefont {Das}, \citenamefont {Dasgupta},\ and\ \citenamefont {Saha-Dasgupta}}]{PhysRevB.102.220502}%
  \BibitemOpen
  \bibfield  {author} {\bibinfo {author} {\bibfnamefont {S.}~\bibnamefont {Bandyopadhyay}}, \bibinfo {author} {\bibfnamefont {P.}~\bibnamefont {Adhikary}}, \bibinfo {author} {\bibfnamefont {T.}~\bibnamefont {Das}}, \bibinfo {author} {\bibfnamefont {I.}~\bibnamefont {Dasgupta}},\ and\ \bibinfo {author} {\bibfnamefont {T.}~\bibnamefont {Saha-Dasgupta}},\ }\bibfield  {title} {\bibinfo {title} {Superconductivity in infinite-layer nickelates: Role of $f$ orbitals},\ }\href {https://doi.org/10.1103/PhysRevB.102.220502} {\bibfield  {journal} {\bibinfo  {journal} {Phys. Rev. B}\ }\textbf {\bibinfo {volume} {102}},\ \bibinfo {pages} {220502} (\bibinfo {year} {2020})}\BibitemShut {NoStop}%
\bibitem [{\citenamefont {Yamamoto}\ and\ \citenamefont {Kageyama}(2013)}]{yamamoto2013hydride}%
  \BibitemOpen
  \bibfield  {author} {\bibinfo {author} {\bibfnamefont {T.}~\bibnamefont {Yamamoto}}\ and\ \bibinfo {author} {\bibfnamefont {H.}~\bibnamefont {Kageyama}},\ }\bibfield  {title} {\bibinfo {title} {{Hydride Reductions of Transition Metal Oxides}},\ }\href {https://doi.org/10.1246/cl.130581} {\bibfield  {journal} {\bibinfo  {journal} {Chemistry Letters}\ }\textbf {\bibinfo {volume} {42}},\ \bibinfo {pages} {946} (\bibinfo {year} {2013})}\BibitemShut {NoStop}%
\bibitem [{\citenamefont {Matsumoto}\ \emph {et~al.}(2019)\citenamefont {Matsumoto}, \citenamefont {Hanzawa}, \citenamefont {Sasase}, \citenamefont {Haindl}, \citenamefont {Katase}, \citenamefont {Hiramatsu},\ and\ \citenamefont {Hosono}}]{Matsumoto2019SmFeAsO}%
  \BibitemOpen
  \bibfield  {author} {\bibinfo {author} {\bibfnamefont {J.}~\bibnamefont {Matsumoto}}, \bibinfo {author} {\bibfnamefont {K.}~\bibnamefont {Hanzawa}}, \bibinfo {author} {\bibfnamefont {M.}~\bibnamefont {Sasase}}, \bibinfo {author} {\bibfnamefont {S.}~\bibnamefont {Haindl}}, \bibinfo {author} {\bibfnamefont {T.}~\bibnamefont {Katase}}, \bibinfo {author} {\bibfnamefont {H.}~\bibnamefont {Hiramatsu}},\ and\ \bibinfo {author} {\bibfnamefont {H.}~\bibnamefont {Hosono}},\ }\bibfield  {title} {\bibinfo {title} {Superconductivity at 48 {K} of heavily hydrogen-doped {SmFeAsO} epitaxial films grown by topotactic chemical reaction using {CaH$_2$}},\ }\href {https://doi.org/10.1103/PhysRevMaterials.3.103401} {\bibfield  {journal} {\bibinfo  {journal} {Physical Review Materials}\ }\textbf {\bibinfo {volume} {3}},\ \bibinfo {pages} {103401} (\bibinfo {year} {2019})}\BibitemShut {NoStop}%
\bibitem [{\citenamefont {Gainza}\ \emph {et~al.}(2023)\citenamefont {Gainza}, \citenamefont {López}, \citenamefont {Serrano-Sánchez}, \citenamefont {Rodrigues}, \citenamefont {Rosa}, \citenamefont {Sobrados}, \citenamefont {Nemes}, \citenamefont {Biskup}, \citenamefont {Fernández-Díaz}, \citenamefont {Martínez},\ and\ \citenamefont {Alonso}}]{GAINZA2023101724}%
  \BibitemOpen
  \bibfield  {author} {\bibinfo {author} {\bibfnamefont {J.}~\bibnamefont {Gainza}}, \bibinfo {author} {\bibfnamefont {C.~A.}\ \bibnamefont {López}}, \bibinfo {author} {\bibfnamefont {F.}~\bibnamefont {Serrano-Sánchez}}, \bibinfo {author} {\bibfnamefont {J.~E.~F.}\ \bibnamefont {Rodrigues}}, \bibinfo {author} {\bibfnamefont {A.~D.}\ \bibnamefont {Rosa}}, \bibinfo {author} {\bibfnamefont {M.~I.}\ \bibnamefont {Sobrados}}, \bibinfo {author} {\bibfnamefont {N.~M.}\ \bibnamefont {Nemes}}, \bibinfo {author} {\bibfnamefont {N.}~\bibnamefont {Biskup}}, \bibinfo {author} {\bibfnamefont {M.~T.}\ \bibnamefont {Fernández-Díaz}}, \bibinfo {author} {\bibfnamefont {J.~L.}\ \bibnamefont {Martínez}},\ and\ \bibinfo {author} {\bibfnamefont {J.~A.}\ \bibnamefont {Alonso}},\ }\bibfield  {title} {\bibinfo {title} {{Evidence of hydrogen content and monovalent Ni oxidation state in non-superconducting bulk anchored infinite-layer nickelates}},\ }\href {https://doi.org/https://doi.org/10.1016/j.xcrp.2023.101724} {\bibfield
  {journal} {\bibinfo  {journal} {Cell Reports Physical Science}\ }\textbf {\bibinfo {volume} {4}},\ \bibinfo {pages} {101724} (\bibinfo {year} {2023})}\BibitemShut {NoStop}%
\bibitem [{\citenamefont {Ding}\ \emph {et~al.}(2023)\citenamefont {Ding}, \citenamefont {Tam}, \citenamefont {Sui}, \citenamefont {Zhao}, \citenamefont {Xu}, \citenamefont {Choi}, \citenamefont {Leng}, \citenamefont {Zhang}, \citenamefont {Wu}, \citenamefont {Xiao}, \citenamefont {Zu}, \citenamefont {Garcia-Fernandez}, \citenamefont {Agrestini}, \citenamefont {Wu}, \citenamefont {Wang}, \citenamefont {Gao}, \citenamefont {Li}, \citenamefont {Huang}, \citenamefont {Zhou},\ and\ \citenamefont {Qiao}}]{Ding2023HydrogenNickelates}%
  \BibitemOpen
  \bibfield  {author} {\bibinfo {author} {\bibfnamefont {X.}~\bibnamefont {Ding}}, \bibinfo {author} {\bibfnamefont {C.~C.}\ \bibnamefont {Tam}}, \bibinfo {author} {\bibfnamefont {X.}~\bibnamefont {Sui}}, \bibinfo {author} {\bibfnamefont {Y.}~\bibnamefont {Zhao}}, \bibinfo {author} {\bibfnamefont {M.}~\bibnamefont {Xu}}, \bibinfo {author} {\bibfnamefont {J.}~\bibnamefont {Choi}}, \bibinfo {author} {\bibfnamefont {H.}~\bibnamefont {Leng}}, \bibinfo {author} {\bibfnamefont {J.}~\bibnamefont {Zhang}}, \bibinfo {author} {\bibfnamefont {M.}~\bibnamefont {Wu}}, \bibinfo {author} {\bibfnamefont {H.}~\bibnamefont {Xiao}}, \bibinfo {author} {\bibfnamefont {X.}~\bibnamefont {Zu}}, \bibinfo {author} {\bibfnamefont {M.}~\bibnamefont {Garcia-Fernandez}}, \bibinfo {author} {\bibfnamefont {S.}~\bibnamefont {Agrestini}}, \bibinfo {author} {\bibfnamefont {X.}~\bibnamefont {Wu}}, \bibinfo {author} {\bibfnamefont {Q.}~\bibnamefont {Wang}}, \bibinfo {author} {\bibfnamefont {P.}~\bibnamefont {Gao}}, \bibinfo {author}
  {\bibfnamefont {S.}~\bibnamefont {Li}}, \bibinfo {author} {\bibfnamefont {B.}~\bibnamefont {Huang}}, \bibinfo {author} {\bibfnamefont {K.-J.}\ \bibnamefont {Zhou}},\ and\ \bibinfo {author} {\bibfnamefont {L.}~\bibnamefont {Qiao}},\ }\bibfield  {title} {\bibinfo {title} {{Critical Role of Hydrogen for Superconductivity in Nickelates}},\ }\href {https://doi.org/10.1038/s41586-022-05657-2} {\bibfield  {journal} {\bibinfo  {journal} {Nature}\ }\textbf {\bibinfo {volume} {615}},\ \bibinfo {pages} {244} (\bibinfo {year} {2023})}\BibitemShut {NoStop}%
\bibitem [{\citenamefont {Balakrishnan}\ \emph {et~al.}(2024)\citenamefont {Balakrishnan}, \citenamefont {Segedin}, \citenamefont {Chow}, \citenamefont {Quarterman}, \citenamefont {Muramoto}, \citenamefont {Surendran}, \citenamefont {Patel}, \citenamefont {LaBollita}, \citenamefont {Pan}, \citenamefont {Song}, \citenamefont {Zhang}, \citenamefont {Baggari}, \citenamefont {Jagadish}, \citenamefont {Shao}, \citenamefont {Goodge}, \citenamefont {Kourkoutis}, \citenamefont {Middey}, \citenamefont {Botana}, \citenamefont {Ravichandran}, \citenamefont {Ariando}, \citenamefont {Mundy},\ and\ \citenamefont {Grutter}}]{Balakrishnan2024HydrogenNickelates}%
  \BibitemOpen
  \bibfield  {author} {\bibinfo {author} {\bibfnamefont {P.~P.}\ \bibnamefont {Balakrishnan}}, \bibinfo {author} {\bibfnamefont {D.~F.}\ \bibnamefont {Segedin}}, \bibinfo {author} {\bibfnamefont {L.~E.}\ \bibnamefont {Chow}}, \bibinfo {author} {\bibfnamefont {P.}~\bibnamefont {Quarterman}}, \bibinfo {author} {\bibfnamefont {S.}~\bibnamefont {Muramoto}}, \bibinfo {author} {\bibfnamefont {M.}~\bibnamefont {Surendran}}, \bibinfo {author} {\bibfnamefont {R.~K.}\ \bibnamefont {Patel}}, \bibinfo {author} {\bibfnamefont {H.}~\bibnamefont {LaBollita}}, \bibinfo {author} {\bibfnamefont {G.~A.}\ \bibnamefont {Pan}}, \bibinfo {author} {\bibfnamefont {Q.}~\bibnamefont {Song}}, \bibinfo {author} {\bibfnamefont {Y.}~\bibnamefont {Zhang}}, \bibinfo {author} {\bibfnamefont {I.~E.}\ \bibnamefont {Baggari}}, \bibinfo {author} {\bibfnamefont {K.}~\bibnamefont {Jagadish}}, \bibinfo {author} {\bibfnamefont {Y.-T.}\ \bibnamefont {Shao}}, \bibinfo {author} {\bibfnamefont {B.~H.}\ \bibnamefont {Goodge}}, \bibinfo {author}
  {\bibfnamefont {L.~F.}\ \bibnamefont {Kourkoutis}}, \bibinfo {author} {\bibfnamefont {S.}~\bibnamefont {Middey}}, \bibinfo {author} {\bibfnamefont {A.~S.}\ \bibnamefont {Botana}}, \bibinfo {author} {\bibfnamefont {J.}~\bibnamefont {Ravichandran}}, \bibinfo {author} {\bibfnamefont {A.}~\bibnamefont {Ariando}}, \bibinfo {author} {\bibfnamefont {J.~A.}\ \bibnamefont {Mundy}},\ and\ \bibinfo {author} {\bibfnamefont {A.~J.}\ \bibnamefont {Grutter}},\ }\bibfield  {title} {\bibinfo {title} {Extensive hydrogen incorporation is not necessary for superconductivity in topotactically reduced nickelates},\ }\href {https://doi.org/10.1038/s41467-024-51479-3} {\bibfield  {journal} {\bibinfo  {journal} {Nature Communications}\ }\textbf {\bibinfo {volume} {15}},\ \bibinfo {pages} {7387} (\bibinfo {year} {2024})}\BibitemShut {NoStop}%
\bibitem [{\citenamefont {Gonzalez}\ \emph {et~al.}(2024)\citenamefont {Gonzalez}, \citenamefont {Ievlev}, \citenamefont {Lee}, \citenamefont {Kim}, \citenamefont {Yu}, \citenamefont {Fowlie},\ and\ \citenamefont {Hwang}}]{Gonzalez2024HydrogenNickelates}%
  \BibitemOpen
  \bibfield  {author} {\bibinfo {author} {\bibfnamefont {M.}~\bibnamefont {Gonzalez}}, \bibinfo {author} {\bibfnamefont {A.}~\bibnamefont {Ievlev}}, \bibinfo {author} {\bibfnamefont {K.}~\bibnamefont {Lee}}, \bibinfo {author} {\bibfnamefont {W.}~\bibnamefont {Kim}}, \bibinfo {author} {\bibfnamefont {Y.}~\bibnamefont {Yu}}, \bibinfo {author} {\bibfnamefont {J.}~\bibnamefont {Fowlie}},\ and\ \bibinfo {author} {\bibfnamefont {H.~Y.}\ \bibnamefont {Hwang}},\ }\bibfield  {title} {\bibinfo {title} {Absence of hydrogen insertion into highly crystalline superconducting infinite layer nickelates},\ }\href {https://doi.org/10.1103/PhysRevMaterials.8.084804} {\bibfield  {journal} {\bibinfo  {journal} {Physical Review Materials}\ }\textbf {\bibinfo {volume} {8}},\ \bibinfo {pages} {084804} (\bibinfo {year} {2024})}\BibitemShut {NoStop}%
\bibitem [{\citenamefont {Lee}\ \emph {et~al.}(2020)\citenamefont {Lee}, \citenamefont {Goodge}, \citenamefont {Li}, \citenamefont {Osada}, \citenamefont {Wang}, \citenamefont {Cui}, \citenamefont {Kourkoutis},\ and\ \citenamefont {Hwang}}]{10.1063/5.0005103}%
  \BibitemOpen
  \bibfield  {author} {\bibinfo {author} {\bibfnamefont {K.}~\bibnamefont {Lee}}, \bibinfo {author} {\bibfnamefont {B.~H.}\ \bibnamefont {Goodge}}, \bibinfo {author} {\bibfnamefont {D.}~\bibnamefont {Li}}, \bibinfo {author} {\bibfnamefont {M.}~\bibnamefont {Osada}}, \bibinfo {author} {\bibfnamefont {B.~Y.}\ \bibnamefont {Wang}}, \bibinfo {author} {\bibfnamefont {Y.}~\bibnamefont {Cui}}, \bibinfo {author} {\bibfnamefont {L.~F.}\ \bibnamefont {Kourkoutis}},\ and\ \bibinfo {author} {\bibfnamefont {H.~Y.}\ \bibnamefont {Hwang}},\ }\bibfield  {title} {\bibinfo {title} {Aspects of the synthesis of thin film superconducting infinite-layer nickelates},\ }\href {https://doi.org/10.1063/5.0005103} {\bibfield  {journal} {\bibinfo  {journal} {APL Materials}\ }\textbf {\bibinfo {volume} {8}},\ \bibinfo {pages} {041107} (\bibinfo {year} {2020})}\BibitemShut {NoStop}%
\bibitem [{\citenamefont {Krieger}\ \emph {et~al.}(2022{\natexlab{a}})\citenamefont {Krieger}, \citenamefont {Martinelli}, \citenamefont {Zeng}, \citenamefont {Chow}, \citenamefont {Kummer}, \citenamefont {Arpaia}, \citenamefont {Sala}, \citenamefont {Brookes}, \citenamefont {Ariando}, \citenamefont {Viart}, \citenamefont {Salluzzo}, \citenamefont {Ghiringhelli},\ and\ \citenamefont {Preziosi}}]{Krieger2022Nickelate}%
  \BibitemOpen
  \bibfield  {author} {\bibinfo {author} {\bibfnamefont {G.}~\bibnamefont {Krieger}}, \bibinfo {author} {\bibfnamefont {L.}~\bibnamefont {Martinelli}}, \bibinfo {author} {\bibfnamefont {S.}~\bibnamefont {Zeng}}, \bibinfo {author} {\bibfnamefont {L.~E.}\ \bibnamefont {Chow}}, \bibinfo {author} {\bibfnamefont {K.}~\bibnamefont {Kummer}}, \bibinfo {author} {\bibfnamefont {R.}~\bibnamefont {Arpaia}}, \bibinfo {author} {\bibfnamefont {M.~M.}\ \bibnamefont {Sala}}, \bibinfo {author} {\bibfnamefont {N.~B.}\ \bibnamefont {Brookes}}, \bibinfo {author} {\bibfnamefont {A.}~\bibnamefont {Ariando}}, \bibinfo {author} {\bibfnamefont {N.}~\bibnamefont {Viart}}, \bibinfo {author} {\bibfnamefont {M.}~\bibnamefont {Salluzzo}}, \bibinfo {author} {\bibfnamefont {G.}~\bibnamefont {Ghiringhelli}},\ and\ \bibinfo {author} {\bibfnamefont {D.}~\bibnamefont {Preziosi}},\ }\bibfield  {title} {\bibinfo {title} {{Charge and Spin Correlations in Infinite-Layer Nickelate Superconductors}},\ }\href
  {https://doi.org/10.1103/PhysRevLett.129.027002} {\bibfield  {journal} {\bibinfo  {journal} {Physical Review Letters}\ }\textbf {\bibinfo {volume} {129}},\ \bibinfo {pages} {027002} (\bibinfo {year} {2022}{\natexlab{a}})}\BibitemShut {NoStop}%
\bibitem [{\citenamefont {Krieger}\ \emph {et~al.}(2022{\natexlab{b}})\citenamefont {Krieger}, \citenamefont {Raji}, \citenamefont {Schlur}, \citenamefont {Versini}, \citenamefont {Bouillet}, \citenamefont {Lenertz}, \citenamefont {Robert}, \citenamefont {Gloter}, \citenamefont {Viart},\ and\ \citenamefont {Preziosi}}]{Krieger_2023}%
  \BibitemOpen
  \bibfield  {author} {\bibinfo {author} {\bibfnamefont {G.}~\bibnamefont {Krieger}}, \bibinfo {author} {\bibfnamefont {A.}~\bibnamefont {Raji}}, \bibinfo {author} {\bibfnamefont {L.}~\bibnamefont {Schlur}}, \bibinfo {author} {\bibfnamefont {G.}~\bibnamefont {Versini}}, \bibinfo {author} {\bibfnamefont {C.}~\bibnamefont {Bouillet}}, \bibinfo {author} {\bibfnamefont {M.}~\bibnamefont {Lenertz}}, \bibinfo {author} {\bibfnamefont {J.}~\bibnamefont {Robert}}, \bibinfo {author} {\bibfnamefont {A.}~\bibnamefont {Gloter}}, \bibinfo {author} {\bibfnamefont {N.}~\bibnamefont {Viart}},\ and\ \bibinfo {author} {\bibfnamefont {D.}~\bibnamefont {Preziosi}},\ }\bibfield  {title} {\bibinfo {title} {Synthesis of infinite-layer nickelates and influence of the capping-layer on magnetotransport},\ }\href {https://doi.org/10.1088/1361-6463/aca54a} {\bibfield  {journal} {\bibinfo  {journal} {Journal of Physics D: Applied Physics}\ }\textbf {\bibinfo {volume} {56}},\ \bibinfo {pages} {024003} (\bibinfo {year}
  {2022}{\natexlab{b}})}\BibitemShut {NoStop}%
\bibitem [{\citenamefont {Parzyck}\ \emph {et~al.}(2024)\citenamefont {Parzyck}, \citenamefont {Anil}, \citenamefont {Wu}, \citenamefont {Goodge}, \citenamefont {Roddy}, \citenamefont {Kourkoutis}, \citenamefont {Schlom},\ and\ \citenamefont {Shen}}]{10.1063/5.0197304}%
  \BibitemOpen
  \bibfield  {author} {\bibinfo {author} {\bibfnamefont {C.~T.}\ \bibnamefont {Parzyck}}, \bibinfo {author} {\bibfnamefont {V.}~\bibnamefont {Anil}}, \bibinfo {author} {\bibfnamefont {Y.}~\bibnamefont {Wu}}, \bibinfo {author} {\bibfnamefont {B.~H.}\ \bibnamefont {Goodge}}, \bibinfo {author} {\bibfnamefont {M.}~\bibnamefont {Roddy}}, \bibinfo {author} {\bibfnamefont {L.~F.}\ \bibnamefont {Kourkoutis}}, \bibinfo {author} {\bibfnamefont {D.~G.}\ \bibnamefont {Schlom}},\ and\ \bibinfo {author} {\bibfnamefont {K.~M.}\ \bibnamefont {Shen}},\ }\bibfield  {title} {\bibinfo {title} {Synthesis of thin film infinite-layer nickelates by atomic hydrogen reduction: Clarifying the role of the capping layer},\ }\href {https://doi.org/10.1063/5.0197304} {\bibfield  {journal} {\bibinfo  {journal} {APL Materials}\ }\textbf {\bibinfo {volume} {12}},\ \bibinfo {pages} {031132} (\bibinfo {year} {2024})}\BibitemShut {NoStop}%
\bibitem [{\citenamefont {Si}\ \emph {et~al.}(2020)\citenamefont {Si}, \citenamefont {Xiao}, \citenamefont {Kaufmann}, \citenamefont {Tomczak}, \citenamefont {Lu}, \citenamefont {Zhong},\ and\ \citenamefont {Held}}]{Si2020}%
  \BibitemOpen
  \bibfield  {author} {\bibinfo {author} {\bibfnamefont {L.}~\bibnamefont {Si}}, \bibinfo {author} {\bibfnamefont {W.}~\bibnamefont {Xiao}}, \bibinfo {author} {\bibfnamefont {J.}~\bibnamefont {Kaufmann}}, \bibinfo {author} {\bibfnamefont {J.~M.}\ \bibnamefont {Tomczak}}, \bibinfo {author} {\bibfnamefont {Y.}~\bibnamefont {Lu}}, \bibinfo {author} {\bibfnamefont {Z.}~\bibnamefont {Zhong}},\ and\ \bibinfo {author} {\bibfnamefont {K.}~\bibnamefont {Held}},\ }\bibfield  {title} {\bibinfo {title} {{Topotactic Hydrogen in Nickelate Superconductors and Akin Infinite-Layer Oxides ABO\(_2\)}},\ }\href {https://doi.org/10.1103/PhysRevLett.124.166402} {\bibfield  {journal} {\bibinfo  {journal} {Physical Review Letters}\ }\textbf {\bibinfo {volume} {124}},\ \bibinfo {pages} {166402} (\bibinfo {year} {2020})}\BibitemShut {NoStop}%
\bibitem [{\citenamefont {Cui}\ \emph {et~al.}(2018)\citenamefont {Cui}, \citenamefont {Zhang}, \citenamefont {Li}, \citenamefont {Lin}, \citenamefont {Zhu}, \citenamefont {Wen}, \citenamefont {Wang}, \citenamefont {Sun}, \citenamefont {Ma}, \citenamefont {Li}, \citenamefont {Gong}, \citenamefont {Xie}, \citenamefont {Gu}, \citenamefont {Li}, \citenamefont {Luo}, \citenamefont {Yu},\ and\ \citenamefont {Yu}}]{CUI201811}%
  \BibitemOpen
  \bibfield  {author} {\bibinfo {author} {\bibfnamefont {Y.}~\bibnamefont {Cui}}, \bibinfo {author} {\bibfnamefont {G.}~\bibnamefont {Zhang}}, \bibinfo {author} {\bibfnamefont {H.}~\bibnamefont {Li}}, \bibinfo {author} {\bibfnamefont {H.}~\bibnamefont {Lin}}, \bibinfo {author} {\bibfnamefont {X.}~\bibnamefont {Zhu}}, \bibinfo {author} {\bibfnamefont {H.-H.}\ \bibnamefont {Wen}}, \bibinfo {author} {\bibfnamefont {G.}~\bibnamefont {Wang}}, \bibinfo {author} {\bibfnamefont {J.}~\bibnamefont {Sun}}, \bibinfo {author} {\bibfnamefont {M.}~\bibnamefont {Ma}}, \bibinfo {author} {\bibfnamefont {Y.}~\bibnamefont {Li}}, \bibinfo {author} {\bibfnamefont {D.}~\bibnamefont {Gong}}, \bibinfo {author} {\bibfnamefont {T.}~\bibnamefont {Xie}}, \bibinfo {author} {\bibfnamefont {Y.}~\bibnamefont {Gu}}, \bibinfo {author} {\bibfnamefont {S.}~\bibnamefont {Li}}, \bibinfo {author} {\bibfnamefont {H.}~\bibnamefont {Luo}}, \bibinfo {author} {\bibfnamefont {P.}~\bibnamefont {Yu}},\ and\ \bibinfo {author} {\bibfnamefont
  {W.}~\bibnamefont {Yu}},\ }\bibfield  {title} {\bibinfo {title} {{Protonation induced high-Tc phases in iron-based superconductors evidenced by NMR and magnetization measurements}},\ }\href {https://doi.org/https://doi.org/10.1016/j.scib.2017.12.009} {\bibfield  {journal} {\bibinfo  {journal} {Science Bulletin}\ }\textbf {\bibinfo {volume} {63}},\ \bibinfo {pages} {11} (\bibinfo {year} {2018})}\BibitemShut {NoStop}%
\bibitem [{\citenamefont {Zeng}\ \emph {et~al.}(2020)\citenamefont {Zeng}, \citenamefont {Tang}, \citenamefont {Yin}, \citenamefont {Li}, \citenamefont {Li}, \citenamefont {Huang}, \citenamefont {Hu}, \citenamefont {Liu}, \citenamefont {Omar}, \citenamefont {Jani}, \citenamefont {Lim}, \citenamefont {Han}, \citenamefont {Wan}, \citenamefont {Yang}, \citenamefont {Pennycook}, \citenamefont {Wee},\ and\ \citenamefont {Ariando}}]{Zeng2020PhaseDiagram}%
  \BibitemOpen
  \bibfield  {author} {\bibinfo {author} {\bibfnamefont {S.}~\bibnamefont {Zeng}}, \bibinfo {author} {\bibfnamefont {C.~S.}\ \bibnamefont {Tang}}, \bibinfo {author} {\bibfnamefont {X.}~\bibnamefont {Yin}}, \bibinfo {author} {\bibfnamefont {C.}~\bibnamefont {Li}}, \bibinfo {author} {\bibfnamefont {M.}~\bibnamefont {Li}}, \bibinfo {author} {\bibfnamefont {Z.}~\bibnamefont {Huang}}, \bibinfo {author} {\bibfnamefont {J.}~\bibnamefont {Hu}}, \bibinfo {author} {\bibfnamefont {W.}~\bibnamefont {Liu}}, \bibinfo {author} {\bibfnamefont {G.~J.}\ \bibnamefont {Omar}}, \bibinfo {author} {\bibfnamefont {H.}~\bibnamefont {Jani}}, \bibinfo {author} {\bibfnamefont {Z.~S.}\ \bibnamefont {Lim}}, \bibinfo {author} {\bibfnamefont {K.}~\bibnamefont {Han}}, \bibinfo {author} {\bibfnamefont {D.}~\bibnamefont {Wan}}, \bibinfo {author} {\bibfnamefont {P.}~\bibnamefont {Yang}}, \bibinfo {author} {\bibfnamefont {S.~J.}\ \bibnamefont {Pennycook}}, \bibinfo {author} {\bibfnamefont {A.~T.~S.}\ \bibnamefont {Wee}},\ and\ \bibinfo {author}
  {\bibfnamefont {A.}~\bibnamefont {Ariando}},\ }\bibfield  {title} {\bibinfo {title} {{Phase Diagram and Superconducting Dome of Infinite-Layer {Nd$_{1-x}$Sr$_x$NiO$_2$} Thin Films}},\ }\href {https://doi.org/10.1103/PhysRevLett.125.147003} {\bibfield  {journal} {\bibinfo  {journal} {Physical Review Letters}\ }\textbf {\bibinfo {volume} {125}},\ \bibinfo {pages} {147003} (\bibinfo {year} {2020})}\BibitemShut {NoStop}%
\bibitem [{\citenamefont {Chow}\ and\ \citenamefont {Ariando}(2022)}]{Chow2022InfiniteLayer}%
  \BibitemOpen
  \bibfield  {author} {\bibinfo {author} {\bibfnamefont {L.~E.}\ \bibnamefont {Chow}}\ and\ \bibinfo {author} {\bibfnamefont {A.}~\bibnamefont {Ariando}},\ }\bibfield  {title} {\bibinfo {title} {{Infinite-Layer Nickelate Superconductors: A Current Experimental Perspective of the Crystal and Electronic Structures}},\ }\href {https://doi.org/10.3389/fphy.2022.834658} {\bibfield  {journal} {\bibinfo  {journal} {Frontiers in Physics}\ }\textbf {\bibinfo {volume} {10}},\ \bibinfo {pages} {834658} (\bibinfo {year} {2022})}\BibitemShut {NoStop}%
\bibitem [{\citenamefont {Li}\ \emph {et~al.}(2020{\natexlab{a}})\citenamefont {Li}, \citenamefont {Wang}, \citenamefont {Lee}, \citenamefont {Harvey}, \citenamefont {Osada}, \citenamefont {Goodge}, \citenamefont {Kourkoutis},\ and\ \citenamefont {Hwang}}]{Li2020SuperconductingDome}%
  \BibitemOpen
  \bibfield  {author} {\bibinfo {author} {\bibfnamefont {D.}~\bibnamefont {Li}}, \bibinfo {author} {\bibfnamefont {B.~Y.}\ \bibnamefont {Wang}}, \bibinfo {author} {\bibfnamefont {K.}~\bibnamefont {Lee}}, \bibinfo {author} {\bibfnamefont {S.~P.}\ \bibnamefont {Harvey}}, \bibinfo {author} {\bibfnamefont {M.}~\bibnamefont {Osada}}, \bibinfo {author} {\bibfnamefont {B.~H.}\ \bibnamefont {Goodge}}, \bibinfo {author} {\bibfnamefont {L.~F.}\ \bibnamefont {Kourkoutis}},\ and\ \bibinfo {author} {\bibfnamefont {H.~Y.}\ \bibnamefont {Hwang}},\ }\bibfield  {title} {\bibinfo {title} {{Superconducting Dome in {Nd$_{1-x}$Sr$_x$NiO$_2$} Infinite Layer Films}},\ }\href {https://doi.org/10.1103/PhysRevLett.125.027001} {\bibfield  {journal} {\bibinfo  {journal} {Physical Review Letters}\ }\textbf {\bibinfo {volume} {125}},\ \bibinfo {pages} {027001} (\bibinfo {year} {2020}{\natexlab{a}})}\BibitemShut {NoStop}%
\bibitem [{\citenamefont {Osada}\ \emph {et~al.}(2022)\citenamefont {Osada}, \citenamefont {Wang}, \citenamefont {Harvey}, \citenamefont {Lee}, \citenamefont {Li}, \citenamefont {Goodge}, \citenamefont {Kourkoutis},\ and\ \citenamefont {Hwang}}]{Osada2022FieldEffect}%
  \BibitemOpen
  \bibfield  {author} {\bibinfo {author} {\bibfnamefont {M.}~\bibnamefont {Osada}}, \bibinfo {author} {\bibfnamefont {B.~Y.}\ \bibnamefont {Wang}}, \bibinfo {author} {\bibfnamefont {S.~P.}\ \bibnamefont {Harvey}}, \bibinfo {author} {\bibfnamefont {K.}~\bibnamefont {Lee}}, \bibinfo {author} {\bibfnamefont {D.}~\bibnamefont {Li}}, \bibinfo {author} {\bibfnamefont {B.~H.}\ \bibnamefont {Goodge}}, \bibinfo {author} {\bibfnamefont {L.~F.}\ \bibnamefont {Kourkoutis}},\ and\ \bibinfo {author} {\bibfnamefont {H.~Y.}\ \bibnamefont {Hwang}},\ }\bibfield  {title} {\bibinfo {title} {{Field-Effect Modulation of Superconductivity in Designer Oxide Superlattices}},\ }\href {https://doi.org/10.1002/adma.202104083} {\bibfield  {journal} {\bibinfo  {journal} {Advanced Materials}\ }\textbf {\bibinfo {volume} {34}},\ \bibinfo {pages} {2104083} (\bibinfo {year} {2022})}\BibitemShut {NoStop}%
\bibitem [{\citenamefont {Perdew}\ \emph {et~al.}(1996)\citenamefont {Perdew}, \citenamefont {Burke},\ and\ \citenamefont {Ernzerhof}}]{PhysRevLett.77.3865}%
  \BibitemOpen
  \bibfield  {author} {\bibinfo {author} {\bibfnamefont {J.~P.}\ \bibnamefont {Perdew}}, \bibinfo {author} {\bibfnamefont {K.}~\bibnamefont {Burke}},\ and\ \bibinfo {author} {\bibfnamefont {M.}~\bibnamefont {Ernzerhof}},\ }\bibfield  {title} {\bibinfo {title} {{Generalized Gradient Approximation Made Simple}},\ }\href {https://doi.org/10.1103/PhysRevLett.77.3865} {\bibfield  {journal} {\bibinfo  {journal} {Phys. Rev. Lett.}\ }\textbf {\bibinfo {volume} {77}},\ \bibinfo {pages} {3865} (\bibinfo {year} {1996})}\BibitemShut {NoStop}%
\bibitem [{\citenamefont {Bl\"ochl}(1994)}]{PhysRevB.50.17953}%
  \BibitemOpen
  \bibfield  {author} {\bibinfo {author} {\bibfnamefont {P.~E.}\ \bibnamefont {Bl\"ochl}},\ }\bibfield  {title} {\bibinfo {title} {Projector augmented-wave method},\ }\href {https://doi.org/10.1103/PhysRevB.50.17953} {\bibfield  {journal} {\bibinfo  {journal} {Phys. Rev. B}\ }\textbf {\bibinfo {volume} {50}},\ \bibinfo {pages} {17953} (\bibinfo {year} {1994})}\BibitemShut {NoStop}%
\bibitem [{\citenamefont {{A.R. Tackett and N.A.W. Holzwarth and G.E. Matthews}}(2001)}]{TACKETT2001348}%
  \BibitemOpen
  \bibfield  {author} {\bibinfo {author} {\bibnamefont {{A.R. Tackett and N.A.W. Holzwarth and G.E. Matthews}}},\ }\bibfield  {title} {\bibinfo {title} {{A Projector Augmented Wave ({PAW}) code for electronic structure calculations, Part {II}: pwpaw for periodic solids in a plane wave basis}},\ }\href {https://doi.org/https://doi.org/10.1016/S0010-4655(00)00241-1} {\bibfield  {journal} {\bibinfo  {journal} {Computer Physics Communications}\ }\textbf {\bibinfo {volume} {135}},\ \bibinfo {pages} {348} (\bibinfo {year} {2001})}\BibitemShut {NoStop}%
\bibitem [{\citenamefont {Paier}\ \emph {et~al.}(2005)\citenamefont {Paier}, \citenamefont {Hirschl}, \citenamefont {Marsman},\ and\ \citenamefont {Kresse}}]{10.1063/1.1926272}%
  \BibitemOpen
  \bibfield  {author} {\bibinfo {author} {\bibfnamefont {J.}~\bibnamefont {Paier}}, \bibinfo {author} {\bibfnamefont {R.}~\bibnamefont {Hirschl}}, \bibinfo {author} {\bibfnamefont {M.}~\bibnamefont {Marsman}},\ and\ \bibinfo {author} {\bibfnamefont {G.}~\bibnamefont {Kresse}},\ }\bibfield  {title} {\bibinfo {title} {{The {P}erdew–{B}urke–{E}rnzerhof exchange-correlation functional applied to the {G}2-1 test set using a plane-wave basis set}},\ }\href {https://doi.org/10.1063/1.1926272} {\bibfield  {journal} {\bibinfo  {journal} {The Journal of Chemical Physics}\ }\textbf {\bibinfo {volume} {122}},\ \bibinfo {pages} {234102} (\bibinfo {year} {2005})}\BibitemShut {NoStop}%
\bibitem [{\citenamefont {Mostofi}\ \emph {et~al.}(2014)\citenamefont {Mostofi}, \citenamefont {Yates}, \citenamefont {Pizzi}, \citenamefont {Lee}, \citenamefont {Souza}, \citenamefont {Vanderbilt},\ and\ \citenamefont {Marzari}}]{MOSTOFI20142309}%
  \BibitemOpen
  \bibfield  {author} {\bibinfo {author} {\bibfnamefont {A.~A.}\ \bibnamefont {Mostofi}}, \bibinfo {author} {\bibfnamefont {J.~R.}\ \bibnamefont {Yates}}, \bibinfo {author} {\bibfnamefont {G.}~\bibnamefont {Pizzi}}, \bibinfo {author} {\bibfnamefont {Y.-S.}\ \bibnamefont {Lee}}, \bibinfo {author} {\bibfnamefont {I.}~\bibnamefont {Souza}}, \bibinfo {author} {\bibfnamefont {D.}~\bibnamefont {Vanderbilt}},\ and\ \bibinfo {author} {\bibfnamefont {N.}~\bibnamefont {Marzari}},\ }\bibfield  {title} {\bibinfo {title} {{An updated version of wannier90: A tool for obtaining maximally-localised Wannier functions}},\ }\href {https://doi.org/https://doi.org/10.1016/j.cpc.2014.05.003} {\bibfield  {journal} {\bibinfo  {journal} {Computer Physics Communications}\ }\textbf {\bibinfo {volume} {185}},\ \bibinfo {pages} {2309} (\bibinfo {year} {2014})}\BibitemShut {NoStop}%
\bibitem [{\citenamefont {Wang}\ \emph {et~al.}(2021)\citenamefont {Wang}, \citenamefont {Xu}, \citenamefont {Liu}, \citenamefont {Tang},\ and\ \citenamefont {Geng}}]{WANG2021108033}%
  \BibitemOpen
  \bibfield  {author} {\bibinfo {author} {\bibfnamefont {V.}~\bibnamefont {Wang}}, \bibinfo {author} {\bibfnamefont {N.}~\bibnamefont {Xu}}, \bibinfo {author} {\bibfnamefont {J.-C.}\ \bibnamefont {Liu}}, \bibinfo {author} {\bibfnamefont {G.}~\bibnamefont {Tang}},\ and\ \bibinfo {author} {\bibfnamefont {W.-T.}\ \bibnamefont {Geng}},\ }\bibfield  {title} {\bibinfo {title} {{VASPKIT: A user-friendly interface facilitating high-throughput computing and analysis using VASP code}},\ }\href {https://doi.org/https://doi.org/10.1016/j.cpc.2021.108033} {\bibfield  {journal} {\bibinfo  {journal} {Computer Physics Communications}\ }\textbf {\bibinfo {volume} {267}},\ \bibinfo {pages} {108033} (\bibinfo {year} {2021})}\BibitemShut {NoStop}%
\bibitem [{\citenamefont {SHIBA}\ and\ \citenamefont {YOKOYAMA}(1987)}]{SHIBA1987264}%
  \BibitemOpen
  \bibfield  {author} {\bibinfo {author} {\bibfnamefont {H.}~\bibnamefont {SHIBA}}\ and\ \bibinfo {author} {\bibfnamefont {H.}~\bibnamefont {YOKOYAMA}},\ }\bibfield  {title} {\bibinfo {title} {{VARIATIONAL MONTE CARLO STUDIES OF HIGHLY CORRELATED ELECTRON SYSTEMS}},\ }in\ \href {https://doi.org/https://doi.org/10.1016/B978-1-4832-2920-1.50070-6} {\emph {\bibinfo {booktitle} {Proceedings of the Yamada Conference XVIII on Superconductivity in Highly Correlated Fermion Systems}}},\ \bibinfo {editor} {edited by\ \bibinfo {editor} {\bibfnamefont {M.}~\bibnamefont {Tachiki}}, \bibinfo {editor} {\bibfnamefont {Y.}~\bibnamefont {Muto}},\ and\ \bibinfo {editor} {\bibfnamefont {S.}~\bibnamefont {Maekawa}}}\ (\bibinfo  {publisher} {Elsevier},\ \bibinfo {year} {1987})\ pp.\ \bibinfo {pages} {264--267}\BibitemShut {NoStop}%
\bibitem [{\citenamefont {Paramekanti}\ \emph {et~al.}(2004)\citenamefont {Paramekanti}, \citenamefont {Randeria},\ and\ \citenamefont {Trivedi}}]{PhysRevB.70.054504}%
  \BibitemOpen
  \bibfield  {author} {\bibinfo {author} {\bibfnamefont {A.}~\bibnamefont {Paramekanti}}, \bibinfo {author} {\bibfnamefont {M.}~\bibnamefont {Randeria}},\ and\ \bibinfo {author} {\bibfnamefont {N.}~\bibnamefont {Trivedi}},\ }\bibfield  {title} {\bibinfo {title} {High-${T}_{c}$ superconductors: A variational theory of the superconducting state},\ }\href {https://doi.org/10.1103/PhysRevB.70.054504} {\bibfield  {journal} {\bibinfo  {journal} {Phys. Rev. B}\ }\textbf {\bibinfo {volume} {70}},\ \bibinfo {pages} {054504} (\bibinfo {year} {2004})}\BibitemShut {NoStop}%
\bibitem [{\citenamefont {Yokoyama}\ and\ \citenamefont {Shiba}(1988)}]{doi:10.1143/JPSJ.57.2482}%
  \BibitemOpen
  \bibfield  {author} {\bibinfo {author} {\bibfnamefont {H.}~\bibnamefont {Yokoyama}}\ and\ \bibinfo {author} {\bibfnamefont {H.}~\bibnamefont {Shiba}},\ }\bibfield  {title} {\bibinfo {title} {{Variational Monte-Carlo Studies of Superconductivity in Strongly Correlated Electron Systems}},\ }\href {https://doi.org/10.1143/JPSJ.57.2482} {\bibfield  {journal} {\bibinfo  {journal} {Journal of the Physical Society of Japan}\ }\textbf {\bibinfo {volume} {57}},\ \bibinfo {pages} {2482} (\bibinfo {year} {1988})}\BibitemShut {NoStop}%
\bibitem [{\citenamefont {Medhi}\ \emph {et~al.}(2007)\citenamefont {Medhi}, \citenamefont {Basu},\ and\ \citenamefont {Kadolkar}}]{PhysRevB.76.235122}%
  \BibitemOpen
  \bibfield  {author} {\bibinfo {author} {\bibfnamefont {A.}~\bibnamefont {Medhi}}, \bibinfo {author} {\bibfnamefont {S.}~\bibnamefont {Basu}},\ and\ \bibinfo {author} {\bibfnamefont {C.~Y.}\ \bibnamefont {Kadolkar}},\ }\bibfield  {title} {\bibinfo {title} {Coexistence of magnetism and superconductivity in a $t\text{\ensuremath{-}}j$ bilayer},\ }\href {https://doi.org/10.1103/PhysRevB.76.235122} {\bibfield  {journal} {\bibinfo  {journal} {Phys. Rev. B}\ }\textbf {\bibinfo {volume} {76}},\ \bibinfo {pages} {235122} (\bibinfo {year} {2007})}\BibitemShut {NoStop}%
\bibitem [{\citenamefont {Tahara}\ and\ \citenamefont {Imada}(2008)}]{doi:10.1143/JPSJ.77.114701}%
  \BibitemOpen
  \bibfield  {author} {\bibinfo {author} {\bibfnamefont {D.}~\bibnamefont {Tahara}}\ and\ \bibinfo {author} {\bibfnamefont {M.}~\bibnamefont {Imada}},\ }\bibfield  {title} {\bibinfo {title} {{Variational Monte Carlo Method Combined with Quantum-Number Projection and Multi-Variable Optimization}},\ }\href {https://doi.org/10.1143/JPSJ.77.114701} {\bibfield  {journal} {\bibinfo  {journal} {Journal of the Physical Society of Japan}\ }\textbf {\bibinfo {volume} {77}},\ \bibinfo {pages} {114701} (\bibinfo {year} {2008})}\BibitemShut {NoStop}%
\bibitem [{\citenamefont {Sorella}(2013)}]{Sorella2013}%
  \BibitemOpen
  \bibfield  {author} {\bibinfo {author} {\bibfnamefont {S.}~\bibnamefont {Sorella}},\ }\bibinfo {title} {{Variational Monte Carlo and Markov Chains for Computational Physics}},\ in\ \href {https://doi.org/10.1007/978-3-642-35106-8_8} {\emph {\bibinfo {booktitle} {Strongly Correlated Systems: Numerical Methods}}},\ \bibinfo {editor} {edited by\ \bibinfo {editor} {\bibfnamefont {A.}~\bibnamefont {Avella}}\ and\ \bibinfo {editor} {\bibfnamefont {F.}~\bibnamefont {Mancini}}}\ (\bibinfo  {publisher} {Springer Berlin Heidelberg},\ \bibinfo {address} {Berlin, Heidelberg},\ \bibinfo {year} {2013})\ pp.\ \bibinfo {pages} {207--236}\BibitemShut {NoStop}%
\bibitem [{\citenamefont {Sarder}\ and\ \citenamefont {Medhi}(2022)}]{Sarder_2022}%
  \BibitemOpen
  \bibfield  {author} {\bibinfo {author} {\bibfnamefont {M.~T.~H.}\ \bibnamefont {Sarder}}\ and\ \bibinfo {author} {\bibfnamefont {A.}~\bibnamefont {Medhi}},\ }\bibfield  {title} {\bibinfo {title} {{Feed-forward neural network based variational wave function for the fermionic Hubbard model in one dimension}},\ }\href {https://doi.org/10.1088/1361-648X/ac7d85} {\bibfield  {journal} {\bibinfo  {journal} {Journal of Physics: Condensed Matter}\ }\textbf {\bibinfo {volume} {34}},\ \bibinfo {pages} {375901} (\bibinfo {year} {2022})}\BibitemShut {NoStop}%
\bibitem [{\citenamefont {Sorella}(2001)}]{PhysRevB.64.024512}%
  \BibitemOpen
  \bibfield  {author} {\bibinfo {author} {\bibfnamefont {S.}~\bibnamefont {Sorella}},\ }\bibfield  {title} {\bibinfo {title} {{Generalized Lanczos algorithm for variational quantum Monte Carlo}},\ }\href {https://doi.org/10.1103/PhysRevB.64.024512} {\bibfield  {journal} {\bibinfo  {journal} {Phys. Rev. B}\ }\textbf {\bibinfo {volume} {64}},\ \bibinfo {pages} {024512} (\bibinfo {year} {2001})}\BibitemShut {NoStop}%
\bibitem [{\citenamefont {{M.A. Hayward and M.J. Rosseinsky}}(2003)}]{HAYWARD2003839}%
  \BibitemOpen
  \bibfield  {author} {\bibinfo {author} {\bibnamefont {{M.A. Hayward and M.J. Rosseinsky}}},\ }\bibfield  {title} {\bibinfo {title} {Synthesis of the infinite layer {Ni(I)} phase {NdNiO$_{2+x}$} by low temperature reduction of {NdNiO$_3$} with sodium hydride},\ }\href {https://doi.org/https://doi.org/10.1016/S1293-2558(03)00111-0} {\bibfield  {journal} {\bibinfo  {journal} {Solid State Sciences}\ }\textbf {\bibinfo {volume} {5}},\ \bibinfo {pages} {839} (\bibinfo {year} {2003})},\ \bibinfo {note} {international Conference on Inorganic Materials 2002}\BibitemShut {NoStop}%
\bibitem [{\citenamefont {Yanagisawa}\ \emph {et~al.}(2002)\citenamefont {Yanagisawa}, \citenamefont {Koike}, \citenamefont {Miyazaki},\ and\ \citenamefont {Yamaji}}]{YANAGISAWA20021379}%
  \BibitemOpen
  \bibfield  {author} {\bibinfo {author} {\bibfnamefont {T.}~\bibnamefont {Yanagisawa}}, \bibinfo {author} {\bibfnamefont {S.}~\bibnamefont {Koike}}, \bibinfo {author} {\bibfnamefont {M.}~\bibnamefont {Miyazaki}},\ and\ \bibinfo {author} {\bibfnamefont {K.}~\bibnamefont {Yamaji}},\ }\bibfield  {title} {\bibinfo {title} {Stripes and superconductivity in high-{Tc} superconductors},\ }\href {https://doi.org/https://doi.org/10.1016/S0022-3697(02)00036-7} {\bibfield  {journal} {\bibinfo  {journal} {Journal of Physics and Chemistry of Solids}\ }\textbf {\bibinfo {volume} {63}},\ \bibinfo {pages} {1379} (\bibinfo {year} {2002})},\ \bibinfo {note} {proceedings of the 8th ISSP International Symposium}\BibitemShut {NoStop}%
\bibitem [{\citenamefont {Watanabe}\ \emph {et~al.}(2021)\citenamefont {Watanabe}, \citenamefont {Shirakawa}, \citenamefont {Seki}, \citenamefont {Sakakibara}, \citenamefont {Kotani}, \citenamefont {Ikeda},\ and\ \citenamefont {Yunoki}}]{PhysRevResearch.3.033157}%
  \BibitemOpen
  \bibfield  {author} {\bibinfo {author} {\bibfnamefont {H.}~\bibnamefont {Watanabe}}, \bibinfo {author} {\bibfnamefont {T.}~\bibnamefont {Shirakawa}}, \bibinfo {author} {\bibfnamefont {K.}~\bibnamefont {Seki}}, \bibinfo {author} {\bibfnamefont {H.}~\bibnamefont {Sakakibara}}, \bibinfo {author} {\bibfnamefont {T.}~\bibnamefont {Kotani}}, \bibinfo {author} {\bibfnamefont {H.}~\bibnamefont {Ikeda}},\ and\ \bibinfo {author} {\bibfnamefont {S.}~\bibnamefont {Yunoki}},\ }\bibfield  {title} {\bibinfo {title} {Unified description of cuprate superconductors using a four-band $d\text{\ensuremath{-}}p$ model},\ }\href {https://doi.org/10.1103/PhysRevResearch.3.033157} {\bibfield  {journal} {\bibinfo  {journal} {Phys. Rev. Res.}\ }\textbf {\bibinfo {volume} {3}},\ \bibinfo {pages} {033157} (\bibinfo {year} {2021})}\BibitemShut {NoStop}%
\bibitem [{\citenamefont {Schmid}\ \emph {et~al.}(2023)\citenamefont {Schmid}, \citenamefont {Mor\'ee}, \citenamefont {Kaneko}, \citenamefont {Yamaji},\ and\ \citenamefont {Imada}}]{PhysRevX.13.041036}%
  \BibitemOpen
  \bibfield  {author} {\bibinfo {author} {\bibfnamefont {M.~T.}\ \bibnamefont {Schmid}}, \bibinfo {author} {\bibfnamefont {J.-B.}\ \bibnamefont {Mor\'ee}}, \bibinfo {author} {\bibfnamefont {R.}~\bibnamefont {Kaneko}}, \bibinfo {author} {\bibfnamefont {Y.}~\bibnamefont {Yamaji}},\ and\ \bibinfo {author} {\bibfnamefont {M.}~\bibnamefont {Imada}},\ }\bibfield  {title} {\bibinfo {title} {{Superconductivity Studied by Solving Ab Initio Low-Energy Effective Hamiltonians for Carrier Doped ${\mathrm{CaCuO}}_{2}$, ${\mathrm{Bi}}_{2}{\mathrm{Sr}}_{2}{\mathrm{CuO}}_{6}$, ${\mathrm{Bi}}_{2}{\mathrm{Sr}}_{2}{\mathrm{CaCu}}_{2}{\mathrm{O}}_{8}$, and ${\mathrm{HgBa}}_{2}{\mathrm{CuO}}_{4}$}},\ }\href {https://doi.org/10.1103/PhysRevX.13.041036} {\bibfield  {journal} {\bibinfo  {journal} {Phys. Rev. X}\ }\textbf {\bibinfo {volume} {13}},\ \bibinfo {pages} {041036} (\bibinfo {year} {2023})}\BibitemShut {NoStop}%
\bibitem [{\citenamefont {Di~Cataldo}\ \emph {et~al.}(2024)\citenamefont {Di~Cataldo}, \citenamefont {Worm}, \citenamefont {Tomczak}, \citenamefont {Si},\ and\ \citenamefont {Held}}]{DiCataldo2024}%
  \BibitemOpen
  \bibfield  {author} {\bibinfo {author} {\bibfnamefont {S.}~\bibnamefont {Di~Cataldo}}, \bibinfo {author} {\bibfnamefont {P.}~\bibnamefont {Worm}}, \bibinfo {author} {\bibfnamefont {J.~M.}\ \bibnamefont {Tomczak}}, \bibinfo {author} {\bibfnamefont {L.}~\bibnamefont {Si}},\ and\ \bibinfo {author} {\bibfnamefont {K.}~\bibnamefont {Held}},\ }\bibfield  {title} {\bibinfo {title} {Unconventional superconductivity without doping in infinite-layer nickelates under pressure},\ }\href {https://doi.org/10.1038/s41467-024-48169-5} {\bibfield  {journal} {\bibinfo  {journal} {Nature Communications}\ }\textbf {\bibinfo {volume} {15}},\ \bibinfo {pages} {3952} (\bibinfo {year} {2024})}\BibitemShut {NoStop}%
\bibitem [{\citenamefont {Hausoel}\ \emph {et~al.}(2025)\citenamefont {Hausoel}, \citenamefont {Cataldo}, \citenamefont {Kitatani}, \citenamefont {Janson},\ and\ \citenamefont {Held}}]{hausoel2025}%
  \BibitemOpen
  \bibfield  {author} {\bibinfo {author} {\bibfnamefont {A.}~\bibnamefont {Hausoel}}, \bibinfo {author} {\bibfnamefont {S.~D.}\ \bibnamefont {Cataldo}}, \bibinfo {author} {\bibfnamefont {M.}~\bibnamefont {Kitatani}}, \bibinfo {author} {\bibfnamefont {O.}~\bibnamefont {Janson}},\ and\ \bibinfo {author} {\bibfnamefont {K.}~\bibnamefont {Held}},\ }\href {https://arxiv.org/abs/2502.12144} {\bibinfo {title} {Superconducting phase diagram of finite-layer nickelates {Nd$_{n+1}$Ni$_n$O$_{2n+2}$}}} (\bibinfo {year} {2025}),\ \Eprint {https://arxiv.org/abs/2502.12144} {arXiv:2502.12144 [cond-mat.str-el]} \BibitemShut {NoStop}%
\bibitem [{\citenamefont {Kang}\ \emph {et~al.}(2023)\citenamefont {Kang}, \citenamefont {Melnick}, \citenamefont {Semon}, \citenamefont {Ryee}, \citenamefont {Han}, \citenamefont {Kotliar},\ and\ \citenamefont {Choi}}]{Kang2023}%
  \BibitemOpen
  \bibfield  {author} {\bibinfo {author} {\bibfnamefont {B.}~\bibnamefont {Kang}}, \bibinfo {author} {\bibfnamefont {C.}~\bibnamefont {Melnick}}, \bibinfo {author} {\bibfnamefont {P.}~\bibnamefont {Semon}}, \bibinfo {author} {\bibfnamefont {S.}~\bibnamefont {Ryee}}, \bibinfo {author} {\bibfnamefont {M.~J.}\ \bibnamefont {Han}}, \bibinfo {author} {\bibfnamefont {G.}~\bibnamefont {Kotliar}},\ and\ \bibinfo {author} {\bibfnamefont {S.}~\bibnamefont {Choi}},\ }\bibfield  {title} {\bibinfo {title} {{Infinite-layer nickelates as Ni-e$_g$ Hund’s metals}},\ }\href {https://doi.org/10.1038/s41535-023-00568-5} {\bibfield  {journal} {\bibinfo  {journal} {npj Quantum Materials}\ }\textbf {\bibinfo {volume} {8}},\ \bibinfo {pages} {35} (\bibinfo {year} {2023})}\BibitemShut {NoStop}%
\bibitem [{\citenamefont {Zhang}\ \emph {et~al.}(2020)\citenamefont {Zhang}, \citenamefont {Yang},\ and\ \citenamefont {Zhang}}]{PhysRevB.101.020501}%
  \BibitemOpen
  \bibfield  {author} {\bibinfo {author} {\bibfnamefont {G.-M.}\ \bibnamefont {Zhang}}, \bibinfo {author} {\bibfnamefont {Y.-f.}\ \bibnamefont {Yang}},\ and\ \bibinfo {author} {\bibfnamefont {F.-C.}\ \bibnamefont {Zhang}},\ }\bibfield  {title} {\bibinfo {title} {{Self-doped Mott insulator for parent compounds of nickelate superconductors}},\ }\href {https://doi.org/10.1103/PhysRevB.101.020501} {\bibfield  {journal} {\bibinfo  {journal} {Phys. Rev. B}\ }\textbf {\bibinfo {volume} {101}},\ \bibinfo {pages} {020501} (\bibinfo {year} {2020})}\BibitemShut {NoStop}%
\bibitem [{\citenamefont {Zhang}\ and\ \citenamefont {Vishwanath}(2020)}]{PhysRevResearch.2.023112}%
  \BibitemOpen
  \bibfield  {author} {\bibinfo {author} {\bibfnamefont {Y.-H.}\ \bibnamefont {Zhang}}\ and\ \bibinfo {author} {\bibfnamefont {A.}~\bibnamefont {Vishwanath}},\ }\bibfield  {title} {\bibinfo {title} {{Type-II $t\text{\ensuremath{-}}J$ model in superconducting nickelate ${\mathrm{Nd}}_{1\ensuremath{-}x}{\mathrm{Sr}}_{x}{\mathrm{NiO}}_{2}$}},\ }\href {https://doi.org/10.1103/PhysRevResearch.2.023112} {\bibfield  {journal} {\bibinfo  {journal} {Phys. Rev. Res.}\ }\textbf {\bibinfo {volume} {2}},\ \bibinfo {pages} {023112} (\bibinfo {year} {2020})}\BibitemShut {NoStop}%
\bibitem [{\citenamefont {Chang}\ \emph {et~al.}(2020)\citenamefont {Chang}, \citenamefont {Zhao},\ and\ \citenamefont {Ding}}]{Chang_2020}%
  \BibitemOpen
  \bibfield  {author} {\bibinfo {author} {\bibfnamefont {J.}~\bibnamefont {Chang}}, \bibinfo {author} {\bibfnamefont {J.}~\bibnamefont {Zhao}},\ and\ \bibinfo {author} {\bibfnamefont {Y.}~\bibnamefont {Ding}},\ }\bibfield  {title} {\bibinfo {title} {{Hund-Heisenberg model in superconducting infinite-layer nickelates}},\ }\href {https://doi.org/10.1140/epjb/e2020-10343-7} {\bibfield  {journal} {\bibinfo  {journal} {The European Physical Journal B}\ }\textbf {\bibinfo {volume} {93}},\ \bibinfo {pages} {220} (\bibinfo {year} {2020})}\BibitemShut {NoStop}%
\bibitem [{\citenamefont {Choubey}\ and\ \citenamefont {Eremin}(2021)}]{PhysRevB.104.144504}%
  \BibitemOpen
  \bibfield  {author} {\bibinfo {author} {\bibfnamefont {P.}~\bibnamefont {Choubey}}\ and\ \bibinfo {author} {\bibfnamefont {I.~M.}\ \bibnamefont {Eremin}},\ }\bibfield  {title} {\bibinfo {title} {Electronic theory for scanning tunneling microscopy spectra in infinite-layer nickelate superconductors},\ }\href {https://doi.org/10.1103/PhysRevB.104.144504} {\bibfield  {journal} {\bibinfo  {journal} {Phys. Rev. B}\ }\textbf {\bibinfo {volume} {104}},\ \bibinfo {pages} {144504} (\bibinfo {year} {2021})}\BibitemShut {NoStop}%
\bibitem [{\citenamefont {Gu}\ \emph {et~al.}(2020)\citenamefont {Gu}, \citenamefont {Li}, \citenamefont {Wan}, \citenamefont {Li}, \citenamefont {Guo}, \citenamefont {Yang}, \citenamefont {Li}, \citenamefont {Zhu}, \citenamefont {Pan}, \citenamefont {Nie},\ and\ \citenamefont {Wen}}]{Gu2020NatComm}%
  \BibitemOpen
  \bibfield  {author} {\bibinfo {author} {\bibfnamefont {Q.}~\bibnamefont {Gu}}, \bibinfo {author} {\bibfnamefont {Y.}~\bibnamefont {Li}}, \bibinfo {author} {\bibfnamefont {S.}~\bibnamefont {Wan}}, \bibinfo {author} {\bibfnamefont {H.}~\bibnamefont {Li}}, \bibinfo {author} {\bibfnamefont {W.}~\bibnamefont {Guo}}, \bibinfo {author} {\bibfnamefont {H.}~\bibnamefont {Yang}}, \bibinfo {author} {\bibfnamefont {Q.}~\bibnamefont {Li}}, \bibinfo {author} {\bibfnamefont {X.}~\bibnamefont {Zhu}}, \bibinfo {author} {\bibfnamefont {X.}~\bibnamefont {Pan}}, \bibinfo {author} {\bibfnamefont {Y.}~\bibnamefont {Nie}},\ and\ \bibinfo {author} {\bibfnamefont {H.-H.}\ \bibnamefont {Wen}},\ }\bibfield  {title} {\bibinfo {title} {Single particle tunneling spectrum of superconducting {Nd$_{1-x}$Sr$_{x}$NiO$_{2}$} thin films},\ }\href {https://doi.org/10.1038/s41467-020-19908-1} {\bibfield  {journal} {\bibinfo  {journal} {Nature Communications}\ }\textbf {\bibinfo {volume} {11}},\ \bibinfo {pages} {6027} (\bibinfo {year}
  {2020})}\BibitemShut {NoStop}%
\bibitem [{\citenamefont {Li}\ \emph {et~al.}(2020{\natexlab{b}})\citenamefont {Li}, \citenamefont {Wang}, \citenamefont {Lee}, \citenamefont {Harvey}, \citenamefont {Osada}, \citenamefont {Goodge}, \citenamefont {Kourkoutis},\ and\ \citenamefont {Hwang}}]{PhysRevLett.125.027001}%
  \BibitemOpen
  \bibfield  {author} {\bibinfo {author} {\bibfnamefont {D.}~\bibnamefont {Li}}, \bibinfo {author} {\bibfnamefont {B.~Y.}\ \bibnamefont {Wang}}, \bibinfo {author} {\bibfnamefont {K.}~\bibnamefont {Lee}}, \bibinfo {author} {\bibfnamefont {S.~P.}\ \bibnamefont {Harvey}}, \bibinfo {author} {\bibfnamefont {M.}~\bibnamefont {Osada}}, \bibinfo {author} {\bibfnamefont {B.~H.}\ \bibnamefont {Goodge}}, \bibinfo {author} {\bibfnamefont {L.~F.}\ \bibnamefont {Kourkoutis}},\ and\ \bibinfo {author} {\bibfnamefont {H.~Y.}\ \bibnamefont {Hwang}},\ }\bibfield  {title} {\bibinfo {title} {{Superconducting Dome in ${\mathrm{Nd}}_{1\ensuremath{-}x}{\mathrm{Sr}}_{x}{\mathrm{NiO}}_{2}$ Infinite Layer Films}},\ }\href {https://doi.org/10.1103/PhysRevLett.125.027001} {\bibfield  {journal} {\bibinfo  {journal} {Phys. Rev. Lett.}\ }\textbf {\bibinfo {volume} {125}},\ \bibinfo {pages} {027001} (\bibinfo {year} {2020}{\natexlab{b}})}\BibitemShut {NoStop}%
\bibitem [{\citenamefont {Capello}\ \emph {et~al.}(2005)\citenamefont {Capello}, \citenamefont {Becca}, \citenamefont {Fabrizio}, \citenamefont {Sorella},\ and\ \citenamefont {Tosatti}}]{PhysRevLett.94.026406}%
  \BibitemOpen
  \bibfield  {author} {\bibinfo {author} {\bibfnamefont {M.}~\bibnamefont {Capello}}, \bibinfo {author} {\bibfnamefont {F.}~\bibnamefont {Becca}}, \bibinfo {author} {\bibfnamefont {M.}~\bibnamefont {Fabrizio}}, \bibinfo {author} {\bibfnamefont {S.}~\bibnamefont {Sorella}},\ and\ \bibinfo {author} {\bibfnamefont {E.}~\bibnamefont {Tosatti}},\ }\bibfield  {title} {\bibinfo {title} {{Variational Description of Mott Insulators}},\ }\href {https://doi.org/10.1103/PhysRevLett.94.026406} {\bibfield  {journal} {\bibinfo  {journal} {Phys. Rev. Lett.}\ }\textbf {\bibinfo {volume} {94}},\ \bibinfo {pages} {026406} (\bibinfo {year} {2005})}\BibitemShut {NoStop}%
\bibitem [{\citenamefont {Yokoyama}\ \emph {et~al.}(2006)\citenamefont {Yokoyama}, \citenamefont {Ogata},\ and\ \citenamefont {Tanaka}}]{doi:10.1143/JPSJ.75.114706}%
  \BibitemOpen
  \bibfield  {author} {\bibinfo {author} {\bibfnamefont {H.}~\bibnamefont {Yokoyama}}, \bibinfo {author} {\bibfnamefont {M.}~\bibnamefont {Ogata}},\ and\ \bibinfo {author} {\bibfnamefont {Y.}~\bibnamefont {Tanaka}},\ }\bibfield  {title} {\bibinfo {title} {{Mott Transitions and $d$-Wave Superconductivity in Half-Filled-Band Hubbard Model on Square Lattice with Geometric Frustration}},\ }\href {https://doi.org/10.1143/JPSJ.75.114706} {\bibfield  {journal} {\bibinfo  {journal} {Journal of the Physical Society of Japan}\ }\textbf {\bibinfo {volume} {75}},\ \bibinfo {pages} {114706} (\bibinfo {year} {2006})}\BibitemShut {NoStop}%
\bibitem [{\citenamefont {De~Franco}\ \emph {et~al.}(2018)\citenamefont {De~Franco}, \citenamefont {Tocchio},\ and\ \citenamefont {Becca}}]{PhysRevB.98.075117}%
  \BibitemOpen
  \bibfield  {author} {\bibinfo {author} {\bibfnamefont {C.}~\bibnamefont {De~Franco}}, \bibinfo {author} {\bibfnamefont {L.~F.}\ \bibnamefont {Tocchio}},\ and\ \bibinfo {author} {\bibfnamefont {F.}~\bibnamefont {Becca}},\ }\bibfield  {title} {\bibinfo {title} {{Metal-insulator transitions, superconductivity, and magnetism in the two-band Hubbard model}},\ }\href {https://doi.org/10.1103/PhysRevB.98.075117} {\bibfield  {journal} {\bibinfo  {journal} {Phys. Rev. B}\ }\textbf {\bibinfo {volume} {98}},\ \bibinfo {pages} {075117} (\bibinfo {year} {2018})}\BibitemShut {NoStop}%
\bibitem [{\citenamefont {V.}\ and\ \citenamefont {Medhi}(2024)}]{PhysRevB.110.125125}%
  \BibitemOpen
  \bibfield  {author} {\bibinfo {author} {\bibfnamefont {K.}~\bibnamefont {V.}}\ and\ \bibinfo {author} {\bibfnamefont {A.}~\bibnamefont {Medhi}},\ }\bibfield  {title} {\bibinfo {title} {{Convolutional restricted Boltzmann machine correlated variational wave function for the Hubbard model on a square lattice}},\ }\href {https://doi.org/10.1103/PhysRevB.110.125125} {\bibfield  {journal} {\bibinfo  {journal} {Phys. Rev. B}\ }\textbf {\bibinfo {volume} {110}},\ \bibinfo {pages} {125125} (\bibinfo {year} {2024})}\BibitemShut {NoStop}%
\bibitem [{\citenamefont {Gutzwiller}(1963)}]{PhysRevLett.10.159}%
  \BibitemOpen
  \bibfield  {author} {\bibinfo {author} {\bibfnamefont {M.~C.}\ \bibnamefont {Gutzwiller}},\ }\bibfield  {title} {\bibinfo {title} {{Effect of Correlation on the Ferromagnetism of Transition Metals}},\ }\href {https://doi.org/10.1103/PhysRevLett.10.159} {\bibfield  {journal} {\bibinfo  {journal} {Phys. Rev. Lett.}\ }\textbf {\bibinfo {volume} {10}},\ \bibinfo {pages} {159} (\bibinfo {year} {1963})}\BibitemShut {NoStop}%
\bibitem [{\citenamefont {Yokoyama}\ and\ \citenamefont {Shiba}(1987)}]{doi:10.1143/JPSJ.56.1490}%
  \BibitemOpen
  \bibfield  {author} {\bibinfo {author} {\bibfnamefont {H.}~\bibnamefont {Yokoyama}}\ and\ \bibinfo {author} {\bibfnamefont {H.}~\bibnamefont {Shiba}},\ }\bibfield  {title} {\bibinfo {title} {{Variational Monte-Carlo Studies of Hubbard Model. I}},\ }\href {https://doi.org/10.1143/JPSJ.56.1490} {\bibfield  {journal} {\bibinfo  {journal} {Journal of the Physical Society of Japan}\ }\textbf {\bibinfo {volume} {56}},\ \bibinfo {pages} {1490} (\bibinfo {year} {1987})}\BibitemShut {NoStop}%
\end{thebibliography}%
\end{document}